\documentclass[twocolumn,reprint,aps,prd,nofootinbib,floatfix,superscriptaddress]{revtex4-1}
\usepackage[utf8]{inputenc}
\usepackage[utf8]{inputenc}
\usepackage{amsmath}
\usepackage{amsfonts}
\usepackage{amssymb}
\usepackage[left=2cm,right=2cm,top=2cm,bottom=2cm]{geometry}
\usepackage{braket}
\usepackage{dsfont}
\usepackage{hyperref}

\usepackage{graphicx}
\usepackage{tikz}
\usetikzlibrary{arrows,decorations.pathmorphing,backgrounds,positioning,fit,petri,shapes}
\usepackage{lipsum}

\newcommand\blfootnote[1]{%
  \begingroup
  \renewcommand\thefootnote{}\footnote{#1}%
  \addtocounter{footnote}{-1}%
  \endgroup
}

\usepackage{multirow}
\begin{document}

\title{The RSD Sorting Hat: Unmixing Radial Scales in Projection}

\begin{abstract}
 Future data sets will enable cross-correlations between redshift space distortions (RSD) and weak lensing (WL). While photometric lensing and clustering cross-correlations have provided some of the tightest cosmological constraints to date, it is not well understood how to optimally perform similar RSD/WL joint analyses in a lossless way. RSD is typically measured in $3D$ redshift space, but WL is inherently a projected signal, making angular statistics a natural choice for the combined analysis. Thus, we determine the amount of RSD information that can be extracted using projected statistics. Specifically we perform a Fisher analysis to forecast constraints and model bias comparing two different Fingers-of-God (FoG) models using both, the $3D$ power spectrum, $P(k, \mu)$, and tomographic $C(\ell)$. We find that because na\"ive tomographic projection mixes large scales with poorly modelled nonlinear radial modes, it does not provide competitive constraints to the $3D$ RSD power spectrum without the model bias becoming unacceptably large. This is true even in the limit of narrow tomographic bins. In light of this we propose a new radial weighting scheme which unmixes radial RSD scales in projection yielding competitive constraints to the $3D$ RSD power spectrum, while keeping the model bias small. This work lays the groundwork for optimal joint analyses of RSD and cosmic shear.

\end{abstract}

\author{Peter L.~Taylor}
\email{peter.taylor@jpl.nasa.gov}
\affiliation{Jet Propulsion Laboratory, California Institute of Technology, 4800 Oak Grove Drive, Pasadena, CA 91109, USA}
\blfootnote{© 2021. All rights reserved.} 
\author{Katarina Markovi\v{c}}
\affiliation{Jet Propulsion Laboratory, California Institute of Technology, 4800 Oak Grove Drive, Pasadena, CA 91109, USA}
\author{Alkistis Pourtsidou}
\affiliation{School of Physics and Astronomy, Queen Mary University of London, Mile End Road, London E1 4NS, UK}
\affiliation{Department of Physics and Astronomy, University of the Western Cape, Cape Town 7535, South Africa}
\author{Eric Huff}
\affiliation{Jet Propulsion Laboratory, California Institute of Technology, 4800 Oak Grove Drive, Pasadena, CA 91109, USA}

\maketitle

\section{Introduction}
Analyses of the RSD power spectrum from spectroscopic surveys have yielded some of the most precise cosmological constraints to date~\cite{Reid:2012sw, Reid:2012sw, Macaulay:2013swa, Beutler:2013yhm, Gil-Marin:2015sqa, Simpson:2015yfa, blake2011wigglez, Macaulay:2013swa, Alam:2020sor, Guzzo:2008ac}. These will further improve as data from Stage IV experiments including Euclid\footnote{\url{http://euclid-ec.org}}~\cite{Blanchard:2019oqi,Laureijs:2011gra}, the Nancy Grace Roman Space Telescope\footnote{\url{https://www.nasa.gov/roman}}~\cite{spergel2015wide} and the Dark Energy Spectroscopic Instrument (DESI)\footnote{\url{https://www.desi.lbl.gov/}}~\cite{Aghamousa:2016zmz}  become available.
\par As a standalone probe RSD provides extremely tight constraints on gravity on cosmological scales and specifically the logarithmic growth function, $f$~\cite{Lahav:1991wc, Linder:2005in}. Measurements of this parameter can distinguish between theories of modified gravity which are indistinguishable from background expansion probes alone~\cite{Huterer:2013xky}. 
\par Yet, small scale radial modes (high $k_\parallel$) are poorly understood and there are large discrepancies between models~\cite{Hamilton:1997zq,Taruya:2010mx, Bernardeau:2001qr, Chen:2020fxs, Scoccimarro:2004tg, Shirasaki:2020bzh}. Thus, when performing parameter inference one typically assumes a nonlinear RSD model and marginalizes over a set of nuisance parameters which govern the behaviour on intermediate nonlinear scales, while removing (or severely down-weighting) scales deep into the nonlinear regime~\cite{Markovic:2019sva}. This is the primary source of modelling uncertainty limiting the constraining power of RSD surveys.
\par Meanwhile, analyses of photometric surveys, using a combination of weak lensing~\cite{Troxel:2017xyo, Hikage:2018qbn, Asgari:2020wuj}, galaxy clustering~\cite{Elvin-Poole:2017xsf} and the cross-correlation between the two~\cite{Abbott:2017wau}, provide complimentary information. Over the next decade, data from Euclid, the Vera Rubin Observatory~\cite{Mandelbaum:2018ouv} and Roman will enable these measurements to achieve unprecedented precision. 
\par By breaking the degeneracy between cosmological and nuisance parameters, the full `$3 \times 2$ point' constraints including galaxy-galaxy lensing are dramatically more precise than those from photometric cosmic shear or clustering alone. This has been shown with existing data~\cite{Abbott:2017wau, Heymans:2020gsg} and it is predicted~\cite{Tutusaus:2020xmc} that for Euclid, the inclusion of the cross-correlation between the two signals improves the dark energy $w_0-w_a$ figure of merit by a factor of five. 
\par By constraining highly uncertain nonlinear RSD models, increasing the signal-to-noise and acting as an important consistency check -- a combined RSD/WL analysis could provide many of the same advantages as a $3 \times 2$ point analysis and yield some of the most competitive constraints on dark energy and modified gravity. The velocity and position of foreground galaxies is correlated with lensing structure, so the WL and RSD signals should be highly covariant provided that 1) the spectroscopic RSD sample lies at a similar redshift to the peak of the lensing kernel and 2) the WL and RSD samples cover overlapping regions of the sky. This will be the case for the DESI luminous red galaxies (LRGs), emission line galaxies (ELGs), the bright galaxy sample (BGS) and Euclid weakly lensed galaxies, and for the spectroscopic / photometric overlap in Roman~\cite{Eifler:2020vvg}. The Euclid spectroscopic sample meanwhile is likely at too high a redshift to cross-correlate with WL.
\par Several studies have previously sought to exploit this covariance. The analysis of~\cite{Lee:2021dou} used weak lensing data to constrain the galaxy bias of the RSD sample. However as the RSD and WL samples were in non-overlapping fields, the data vector in this study contained no RSD/WL cross-correlation terms. Instead a photometric sample was constructed to match the properties of the RSD sample and the resulting galaxy-galaxy lensing measurement was used to constrain the galaxy bias. Other analyses have also sought to exploit the cross-correlation between photometric cosmic shear and spectroscopic galaxy clustering by performing $3 \times 2$ point analysis~\cite{heymans2020, Joudaki:2017zdt}. Since these studies use photometric redshifts for the lensed galaxies and the tomographic bin widths were taken to be much larger than the nonlinear RSD scale, one should not expect this type of analysis to provide additional constraints on nonlinear RSD i.e the high $k_\parallel$-modes. In fact binning the spectroscopic clustering sample into broad redshift bins leads to a loss of information. But if one had spectral information for the lensed galaxies from cross-matching the RSD and WL catalogs, it is possible that by providing additional measurements of the small scale radial modes, the WL/RSD cross-signal could help constrain nonlinear RSD.
\par Given the large overlap between Euclid and DESI and the Roman photometric and spectroscopic survey~\cite{Eifler:2020vvg}, we anticipate a sample of approximately $30$ million galaxies where we have both lensing and spectroscopic information within the next decade.\footnote{There are also proposals to perform kinematic weak lensing studies which will provide spectra for lensed galaxies~\cite{Huff:2013dha}. A first detection of this signal was recently made in~\cite{gurri2020first}.}  To take advantage of this cross-probe signal, we first need an appropriate two-point estimator.
\par While RSD studies typically use the 3D anisotropic power spectrum\footnote{We will use $P(k,\mu)$ and $P(k_\parallel,k_\perp)$ interchangeably through the remainder of the text using the relations $k_\parallel = k \mu$ and $k = \sqrt{k_\parallel ^ 2 + k_\perp ^2}$.}, $P(k,\mu)$, the anisotropic two-point correlation function, $\xi(s_\perp, s_\parallel)$, or Legendre multipoles $P_\ell (k)$, weak lensing is a projected observable, so cosmic shear studies use angular statistics, in particular the tomographic cosmic shear power spectrum, $C(\ell)$, or the two-point correlation functions, $\xi_{\pm} (\theta)$. Projection is not a completely invertible operation, so we should seek to perform the joint analysis in projected space.
\par This raises the question of whether it is even possible to capture all the information in the anisotropic power spectrum, $P(k,\mu)$ using projected statistics. One may na\"ively wish to use classical projected tomographic power spectra~\cite{Jalilvand:2019brk, Gebhardt:2020imr, Camera:2018jys, Asorey:2012rd, Padmanabhan:2006cia,Loureiro:2018qva}, $C_{ij}(\ell)$, binning galaxies in sufficiently narrow redshift bins (with bin numbers $i$ and $j$) to capture the $k_\parallel$-modes down to sufficiently small scales. As shown in~\cite{Jalilvand:2019brk}, this mixes large and small $k_\parallel$-modes for each fixed perpendicular scale, $k_\perp$. This has two problems:
\begin{itemize}
\item Cosmological constraining power is lost because each ${\bf k} = (k_\parallel, \left| {\bf k_\perp} \right|)$ contains independent information.
\item As we are mixing $k_\parallel$-scales for fixed $\ell$, we must cut data points to avoid model bias from nonlinear RSD scales, even when these data points also contain useful information about linear $k_\parallel$-scales.
\end{itemize}
\par To sidestep these issues, we will show how to construct a radial weighting function (basis) which sorts $k_\parallel$-scales in projection. For radial weights $w(\eta_a,z)$ and $w(\eta_b,z)$ in this new basis, we write the angular power spectrum as $C^{\eta_a \eta_b} (\ell)$. We will demonstrate there are tight correspondences between $\eta-k_\parallel$ and $\ell - k_\perp$. This can be thought of as the RSD analogue of $k$-cut and $x$-cut cosmic shear which sorts lensing 2-point statistics by sensitivity to difference structure scales~\cite{Taylor:2020zcg, Taylor:2018snp, Taylor:2020imc, Bernardeau:2013rda}.

\par We will show that one must still use the RSD power spectrum, $P(k,\mu)$, to optimally extract information from large linear radial scales before discussing how to use the two estimators, $C^{\eta_b \eta_b}(\ell)$ and $P(k,\mu)$, in a joint analysis. This is somewhat similar to the hybrid estimator developed in~\cite{Wang:2020wsx} except we work in projected space on small scales while~\cite{Wang:2020wsx} worked in spherical-Bessel projected space on large scales to handle wide-angle effects. Although this means in the future we will not be able to extract information from large scale radial modes in the WL/RSD cross-signal, this is acceptable as we are primarily interested in the cross-correlations on smaller radial scales to constrain nonlinear RSD.
\par The structure of this paper is as follows. In Section~\ref{sec:formalism} we summarize the formalism of projected RSD before presenting our new basis. In Section~\ref{sec:fisher results} we perform a Fisher analysis. Using two phenomological FoG models, we compare the precision and model bias when using the RSD power spectrum, $P(k,\mu)$, standard tomography, $C_{ij}(\ell)$ and our new basis, $C^{\eta_a \eta_b}(\ell)$, before concluding in Section~\ref{sec:conclusion}. Finally we discuss issues related to performing a likelihood analysis with real data in the Appendix. Throughout the remainder of the paper we take $\Omega_m = 0.315$, $\Omega_b = 0.04$, $h_0 = 0.67$, $n_s = 0.96$ and $\sigma _8 = 0.8$.

\section{RSD Angular Power Spectra} \label{sec:formalism}

\subsection{A Brief Review of the RSD Anisotropic Power Spectrum}
Before discussing projected RSD, we first review the formalism of the anisotropic RSD power spectrum, $P(k,\mu)$. We decompose the RSD spectrum into isotropic and anistropic parts~\cite{Gebhardt:2020imr}
\begin{equation} \label{eq:pkmu}
P(k, \mu) =\widetilde{A}^2_{\mathrm{RSD}}(\mu, k \mu) P(k),
\end{equation}
where $\mu = cos(\theta)$, $\theta$ is the angle between the wave-vector and the line of sight,  $P(k)$ is the matter power spectrum and $\widetilde{A}$ is the RSD operator which takes the power spectrum into redshift space and accounts for galaxy bias. As in~\cite{Gebhardt:2020imr}, we decompose it as
\begin{equation}
\widetilde{A}_{\mathrm{RSD}}(\mu, k \mu)= b_g \left(1+\beta \mu^{2}\right) \widetilde{A}_{\mathrm{nl}}(\mu, k \mu),
\end{equation}
where $b_g \left(1+\beta \mu^{2}\right)$ is the Kaiser term~\cite{kaiser1987clustering} acting on linear scales, $b_g$ is the linear galaxy bias, $\beta = f/b_g$ and $\widetilde{A}_{\mathrm{nl}}$ is the operator that accounts for the nonlinear RSD model.
In this paper we consider two simple phenomenological nonlinear models which account for the FoG. For a Gaussian FoG, the nonlinear RSD operator is~\cite{Hamilton:1997zq}
\begin{equation} \label{eqn:gauss FoG}
\widetilde{A}_{\rm Gauss} (k \mu)=\exp \left[-f^2 \sigma_{v}^{2} k^{2} \mu^{2} \right],
\end{equation}
while for the Lorentzian model it is given by
\begin{equation} \label{eqn:lorentz fog}
\widetilde{A}_{\rm Lorentz}(k \mu)=\frac{1}{1+ f^2 \sigma_{v}^{2} k^{2} \mu^{2}}.
\end{equation}
\par Throughout the remainder of this work we use the nonlinear power spectrum from {\tt Halofit}~\cite{Takahashi:2012em} to generate the isotropic part, $P(k)$ and take the Lorentzian FoG as the fiducial model\footnote{In this work we are primarily interested in the difference between the two models as proxy for nonlinear RSD model uncertainty.} unless explicitly stated otherwise. This phenomenological model is as accurate as more physically motivated models~\cite{Jalilvand:2019brk}. We choose to ignore the impact of spectroscopic redshift uncertainty as this is subdominant to the FoG.

\subsection{Survey Selection Functions, Windows and Tomographic Bins}
\par In this section we aim to clarify some definitions which will be used through the remainder of the text. Given a spectroscopic survey, the observed radial distribution of galaxies will have some distribution in redshift, $z$, which we will refer to as the survey radial selection function and which we write as $N(z)$.
\par We divide the radial selection function into several coarse redshift bins which we will refer to as radial windows or simply as windows. We use the notation $W(z)$ to denote the galaxy redshift distribution of a window. It is standard practice in RSD anisotropic power spectrum studies to sub-divide the survey selection function into coarse redshift bins, assume there is no redshift evolution of the power spectrum in the window and analyse each redshift bin separately. For the remainder of this work we compute distances and the power spectrum at the average redshift of the window $z_w = \int {\rm d} z \ zW(z) / \int {\rm d} z \ W(z)$, writing $r_w$ for the comoving distance at $z_w$. The impact of choosing an effective redshift are studied in~\cite{Obuljen:2021ryv}. We will also restrict our attention to a single window for the remainder of the paper. Finally when we consider projected tomographic power spectra we will further sub-divide the window into tomographic bins in redshift. In this case we denote the $i^{\rm th}$ tomographic bin window as $W_i(z)$.

\subsection{Angular Tomographic Power Spectra in the Flat-Sky Approximation}

\par Since the RSD power spectrum $P(k,\mu)$ is anisotropic, care must be taken when projecting to angular statistics. For the remainder of this work we will assume the plane-parallel approximation~\cite{Gebhardt:2020imr, Jalilvand:2019brk, Matthewson:2020rdt} which is an extremely good approximation for $\ell > 100$ assuming that the radial separation of tomographic bins is small~\cite{Matthewson:2020rdt}. Then for tomographic bin numbers $i$ and $j$, following the notation of~\cite{Gebhardt:2020imr}, we can write the projected tomographic angular power spectrum\footnote{Formally since the RSD field is anisotropic $C_{\ell \ell'} \neq 0$ for $\ell \neq \ell'$. We choose to ignore this effect since the coupling is weak and contained to a few adjacent $\ell$-modes~\cite{Gebhardt:2017chz}. In practice with real data, one must also use wide bandpowers in $\ell$ to deconvolve the survey mask, mitigating the impact of coupling.} as
\begin{equation} \label{eqn:tomo}
C_{ij} (\ell) = \frac{1}{r_ir_j} C \left(k_\perp = \frac{\ell + 1/2}{r_0} ; r_i, r_j\right),
\end{equation}
where $r_i$ and $r_j$ give the comoving radial distance to tomographic bins $i$ and $j$, $r_0 = (r_i + r_j)/2$ and
\begin{equation} \label{eqn:c1}
\begin{aligned}
C \left(k_\perp; r_i, r_j \right) = \frac{1}{\pi}\int_0^\infty {\rm d} k_\parallel \ P(k, \mu)  \widetilde K_{ij} (k_\parallel),
\end{aligned}
\end{equation}
where $k = \sqrt{k_\perp^2 + k_\parallel ^2}$ and $\mu =  k_\parallel / k$, and we define the {\it radial-mode efficiency kernel}, $\widetilde K_{ij} (k_\parallel)$, as
\begin{equation} \label{eq:rad eff}
    \widetilde K_{ij} (k_\parallel) = \widetilde W_i^*(k_\parallel) \widetilde W_j(k_\parallel).
\end{equation}
Here $\widetilde W_i (k_\parallel)$ is the Fourier transform of the radial window for tomographic bin $i$, so that,
\begin{equation} \label{eqn:fourier 1}
    \widetilde W_i(k_\parallel) = \int_{r[z_{\rm min}]}^{r[z_{\rm max}]} {\rm d} r \ W_i\left(r[z] \right) \exp \left( -i k_\parallel r[z] \right)
\end{equation}
with the normalization
\begin{equation}
\int_{r[z_{\rm min}]}^{r[z_{\rm max}]} {\rm d} r \ \ W_i\left(r[z] \right) = 1.
\end{equation}
The radial-mode efficiency kernel, $\widetilde K_{ij} (k_\parallel)$, defined in Eqn.~\ref{eq:rad eff}, is an important quantity. We will refer to it at many points in the remainder of this work as it defines how sensitive the projected spectra are to different $k_\parallel$-modes. It is important to note that it depends on the comoving distance, $r$, and hence is a function of the background cosmology.
\par In the absence of a survey mask, the expected observed power spectrum is a sum of the signal in Eqn.~\ref{eqn:tomo} and the shot-noise. The shot-noise is strictly a function of the number of galaxies in the bin, $N_{i,{\rm gals}}$, so we can use the expression from the isotropic case~\cite{Blanchard:2019oqi}
\begin{equation}
    N_{ij} (\ell) =\frac{\delta_{ij}}{N_{i,{\rm gals}}},
\end{equation}
where $\delta_{ij}$ is the Kronecker delta.

\subsection{Generalized Angular Power Spectra in the Flat-Sky Approximation}
\par To extract information we weight every galaxy by some radial weight function, $w(\eta, z)$. Then inside each window the cosmological information is contained in the {\it generalized power spectra} $C^{\eta_a, \eta_b} (\ell)$. Power spectra with generalized weights have previously been proposed for weak lensing~\cite{Taylor:2018nrc, Taylor:2018qda}. It should be noted that standard tomography is just a special case of the generalized power spectra where one takes $\{ w(\eta, z) \}$ to be a set of narrow non-overlapping top-hat functions.
\par After applying the weighting, $w(\eta, z)$, the radial window is
\begin{equation} \label{eqn:w1}
    \mathcal{W}(\eta,r[z]) = w(\eta,z) W(r[z]).
\end{equation}
Generalizing the expressions in the previous section, it follows that
\begin{equation} \label{eqn:c2a}
C^{\eta_a \eta_b} (\ell) = \frac{1}{r_w^2}C^{\eta_a \eta_b} \left(k_\perp = \frac{\ell + 1/2}{r_w} ; r_w \right)
\end{equation}
where 
\begin{equation} \label{eqn:c2b}
\begin{aligned}
 C^{\eta_a \eta_b} &\left(k_\perp; r_w \right) = \frac{1}{ \pi}\int_0^\infty {\rm d} k_\parallel \ P(k, \mu) \widetilde K(k_\parallel;\eta_a,\eta_b),
\end{aligned}
\end{equation}
and the {\it generalized radial-mode efficiency kernel}, $K(k_\parallel;\eta_a,\eta_b)$, is given by
\begin{equation} \label{eqn:K1}
   \widetilde K(k_\parallel;\eta_a,\eta_b) = \widetilde W^*(\eta_a, k_\parallel)  \widetilde W(\eta_b, k_\parallel).
\end{equation}
The Fourier transform of the radial windows in the new basis are
\begin{equation} \label{eq:fourier window}
    \widetilde W(\eta, k_\parallel) = \int_{r[z_{\rm min}]}^{r[z_{\rm max}]} {\rm d} r  \ \mathcal{W}\left(\eta, r[z] \right) \exp \left( -i k_\parallel r[z] \right).
\end{equation}
Finally the shot-noise is given by
\begin{equation}
   N^{\eta_a \eta_b} (\ell) = \frac{1}{N_{{\rm gals}}} \int_{z_{\rm min}}^{z_{\rm max}}{d} z \ W(z) w(\eta_a, z) w(\eta_b,z),
\end{equation}
where in this case we assume the normalization
\begin{equation}
    \int_{z_{\rm min}}^{z_{\rm max}}{d} z \ W(z) = 1.
\end{equation}
We refer the reader to~\cite{{Heavens:2003jx}} for the derivation of the shot-noise with a generic weighting function. In practice the shot-noise can be estimated from random realizations.

\subsection{The Tomographic Radial-Mode Efficiency Kernel for Top-hat Window Functions}
\par It is in general difficult to compute the radial-mode efficiency kernels at each point in cosmological parameter space rapidly enough for cosmological parameter inference if the window (and hence tomographic bin windows) are not simple analytic functions. We discuss a fast numerical approach\footnote{In practice all results in this paper are computed using this fast numerical approaches given in Appendices~\ref{sec: tomo app}-\ref{sec:app b}.} in Appendix~\ref{sec: tomo app} for more realistic windows. For the time being, we restrict our attention to top-hat tomographic bin windows, $W_i(z)$, since it is possible to proceed analytically. This is a good starting point since in the limit of narrow tomographic bins, the tomographic bin windows are well approximated by top-hat functions.
\par Assuming the tomographic bin window is centered at $r_i$, with bin width $\Delta r_i$, noting that the Fourier transform of a top-hat function is a sinc function, we find~\cite{Gebhardt:2020imr, Jalilvand:2019brk}
 \begin{equation} \label{eqn:Kijtomo}
    \widetilde K_{ii}(k_\parallel) ={\rm sinc}^2 \left(\frac{k_\parallel \Delta r_i}{2} \right),
\end{equation}
where  ${\rm sinc}(x) = \sin(x) /x$.

\begin{figure}[!hbt]
\includegraphics[width = \linewidth]{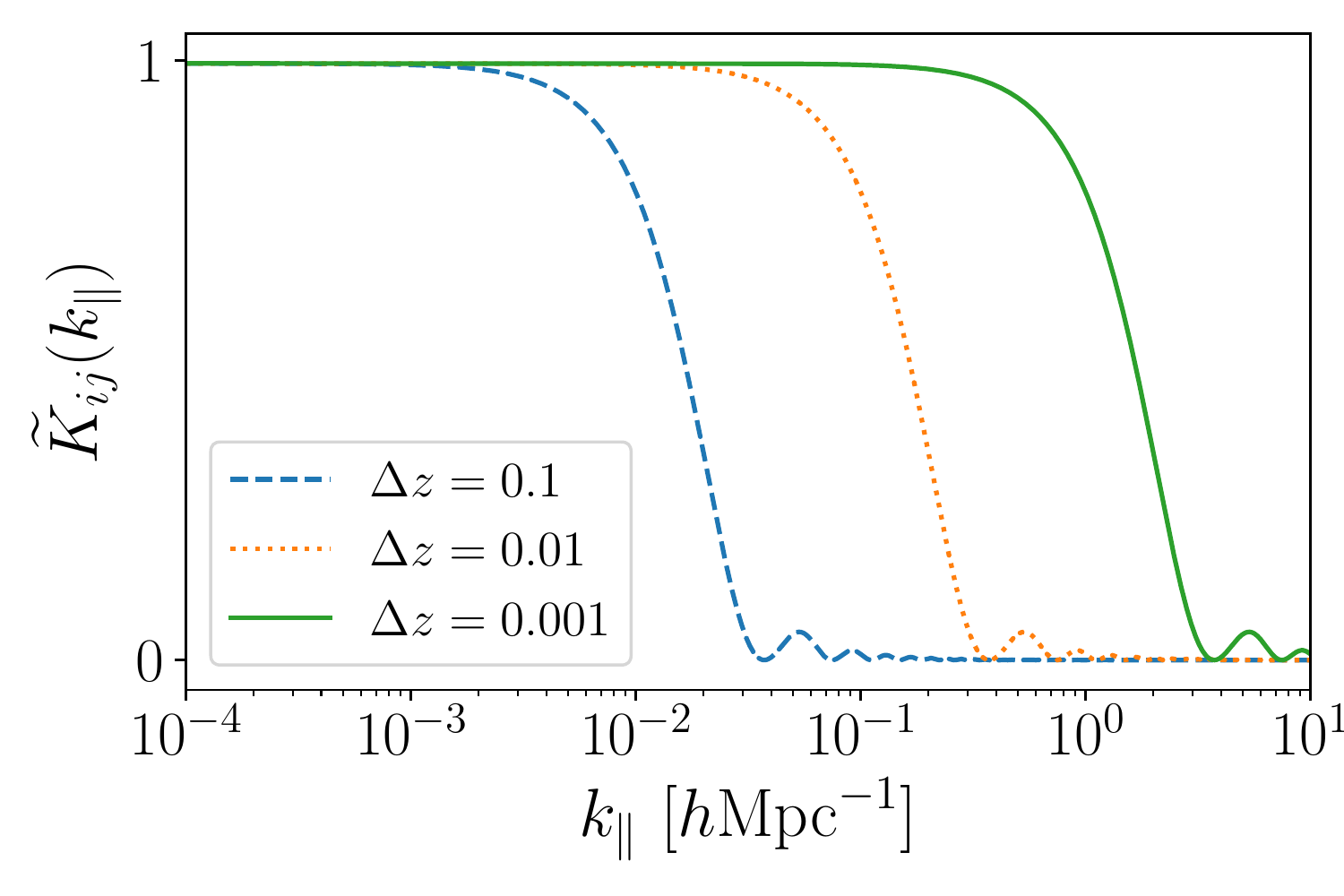}
\caption{The radial mode-efficiency kernels, $\widetilde K_{ij}(k_\parallel)$, for top-hat tomographic windows with redshift widths $\Delta z$, centered around $z=0.5$. The radial mode-efficiency kernel defines the sensitivity of the tomographic $C(\ell)$ to $k_\parallel$-modes in the RSD power spectrum (see Eqn.~\ref{eqn:c1}). Smaller bins widths lead to more sensitivity to smaller scale parallel modes. Since the kernels are broad in $k_\parallel$, even in the limit of small bin widths, information is still lost when using tomographic $C(\ell)$ as we mix independent $k_\parallel$-modes across a broad range of scales.}
\label{fig:tomo_kernel}
\end{figure}

\begin{figure}[!hbt]
\includegraphics[width = \linewidth]{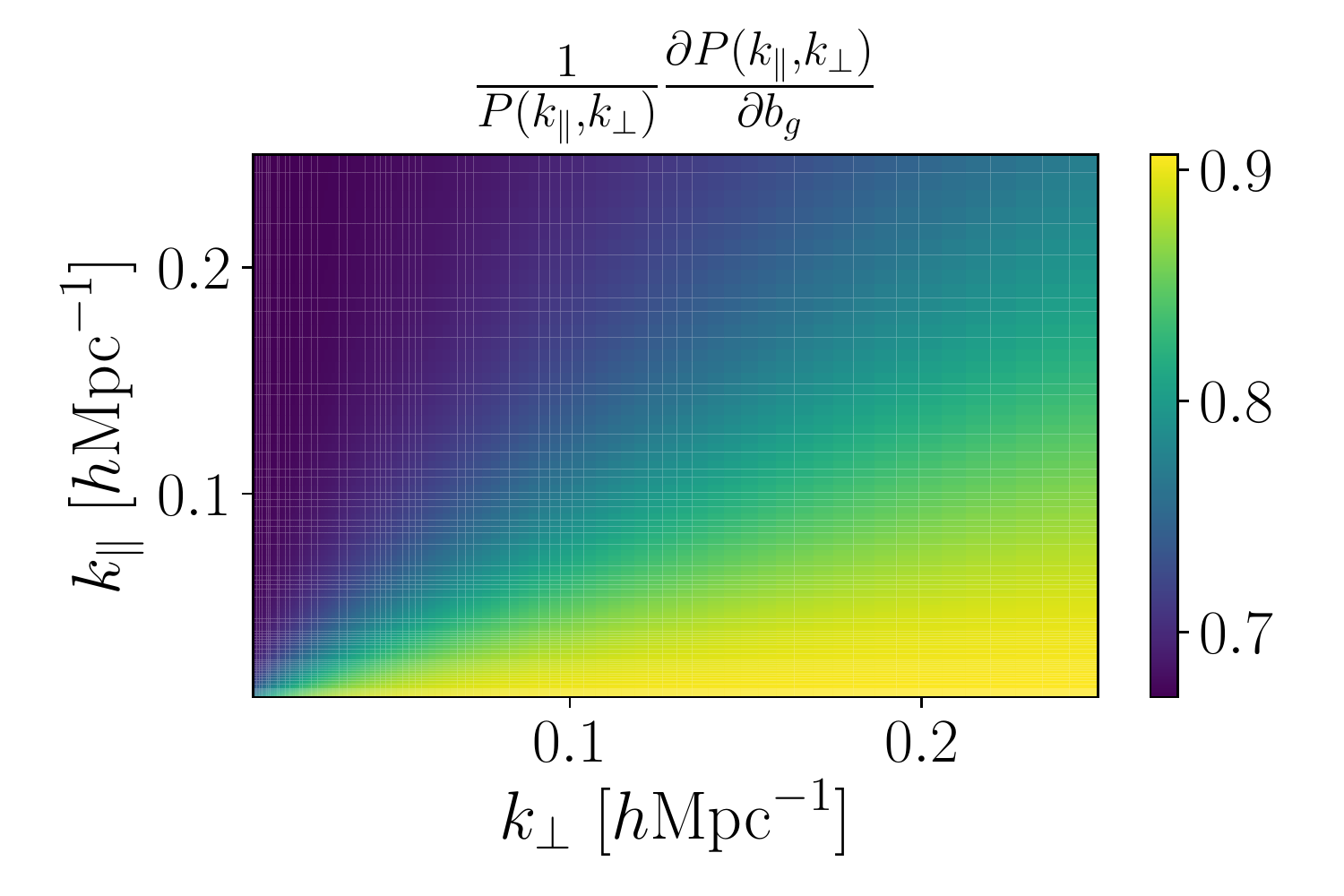}
\includegraphics[width = \linewidth]{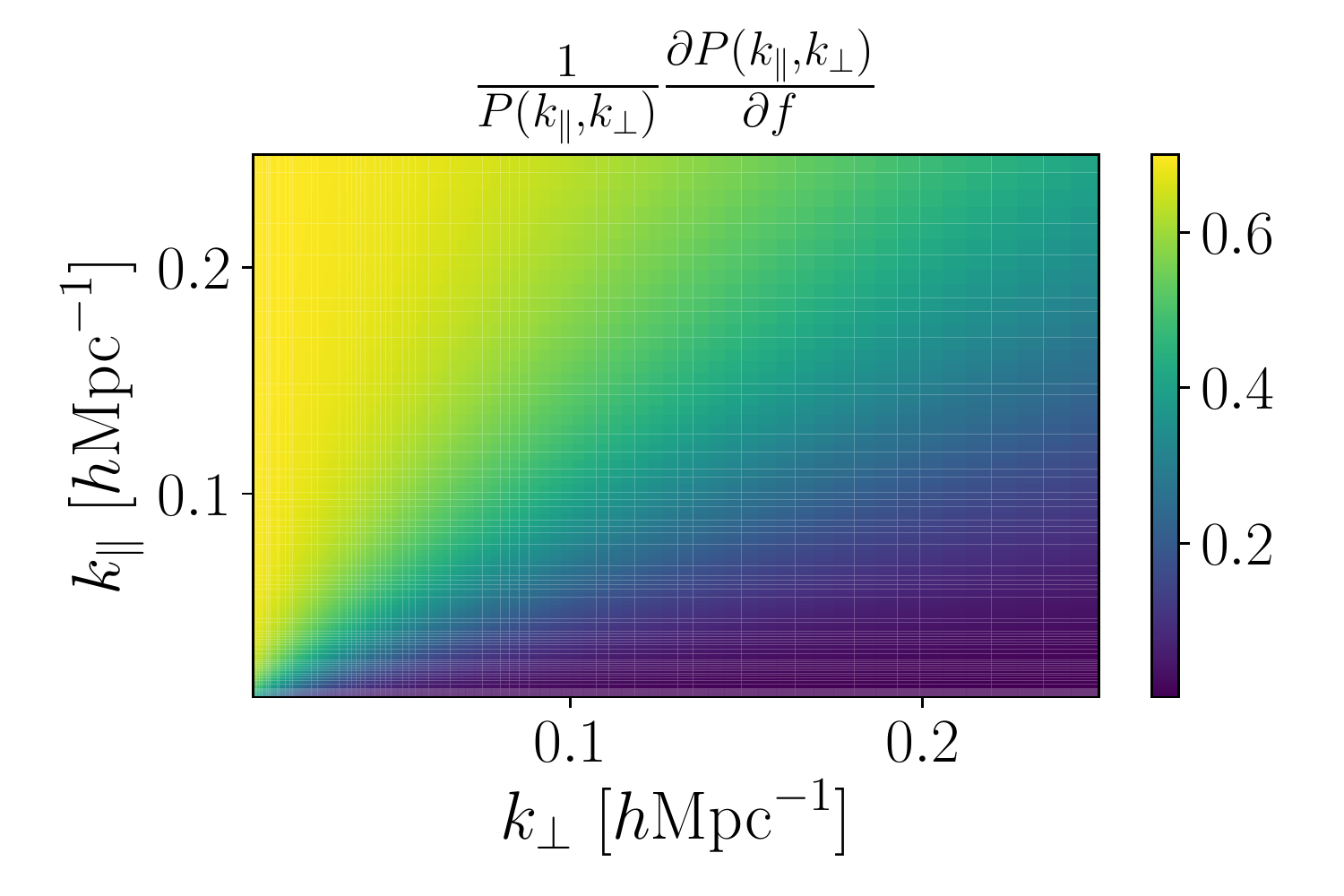}
\includegraphics[width = \linewidth]{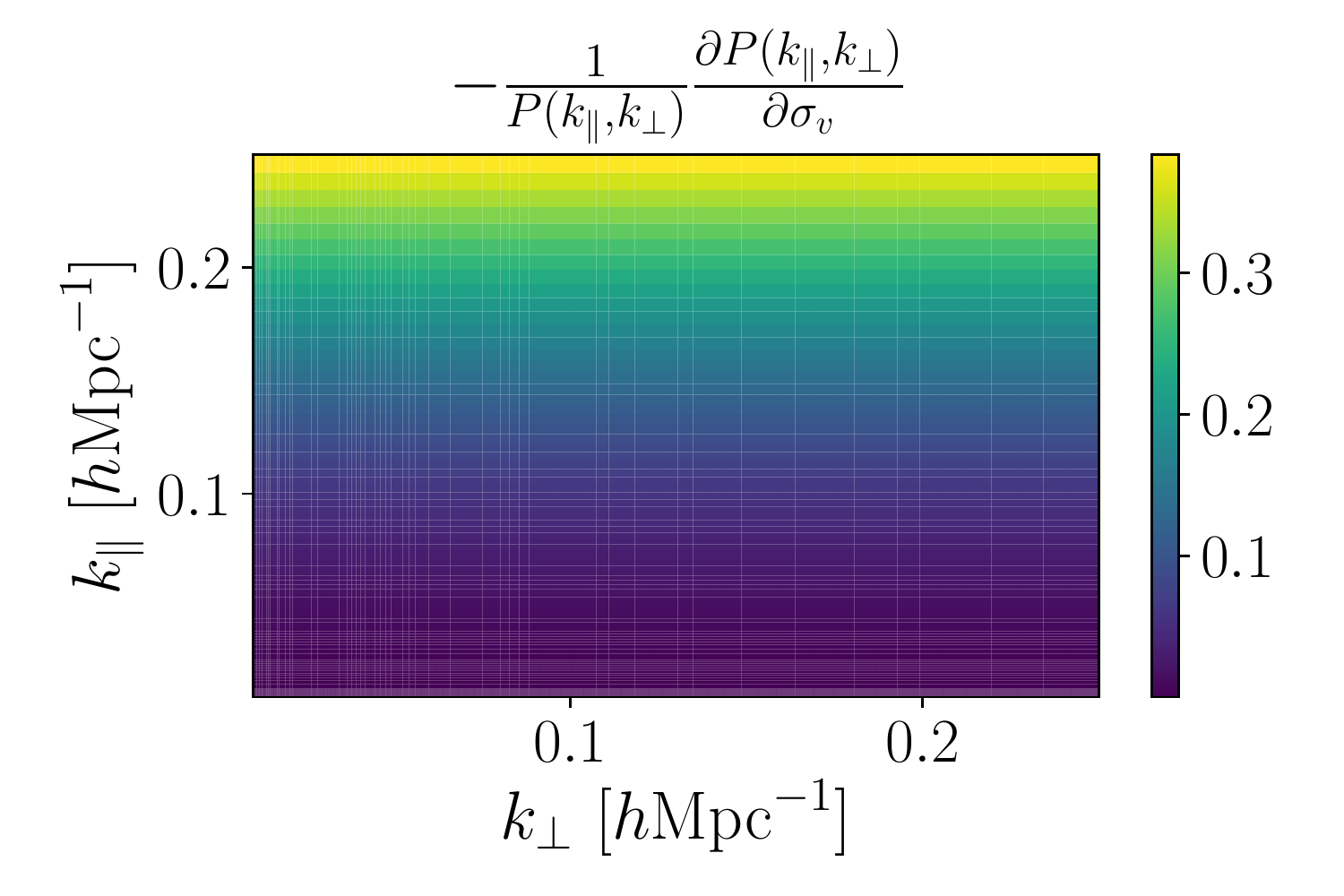}
\caption{The derivative of the RSD power spectrum $P(k_\parallel, k_\perp)$ with respect to $b_g$, $f$ and $\sigma_v$. Each parameter affects the power spectrum in different regions of $k_\parallel - k_\perp$ space. This information is partially lost when integrating over $k_\parallel$ to compute tomographic $C(\ell)s$ (see Eqn.~\ref{eqn:c1}). We quantify this statement in Section~\ref{sec:fisher results}.}
\label{fig:derivatives}
\end{figure}

\par In Fig.~\ref{fig:tomo_kernel} we plot the radial-mode efficiency kernels for tomographic top-hat windows of different widths, $\Delta z$, at a redshift $z=0.5$. Taking the bin width to be $\Delta z = 0.1$, we see that almost all nonlinear RSD information is lost as the radial mode-efficiency kernel is effectively zero for $k_\parallel \gtrsim 0.03 \ h {\rm Mpc}^{-1}$. Decreasing the bin width to $\Delta z = 0.01$, we start to probe the mildly nonlinear regime $k_\parallel \lesssim  0.2 \ h {\rm Mpc}^{-1}$. But if we decrease the bin width to $\Delta z = 0.001$ the tomographic power spectra will remain sensitive to radial modes extremely deep into the nonlinear regime, up to $k_\parallel \lesssim  3 \ h {\rm Mpc}^{-1}$. Hence we see that decreasing the bin width yields sensitivity to larger $k_\parallel$-modes. We will restrict our attention to $\Delta z > 0.01$ to avoid sensitivity to the deeply nonlinear regime for the remainder of this work.
\par It is worth emphasizing two key points. First, tomographic RSD angular power spectra  probe nonlinear radial modes even in the limit of low-$\ell$  (see Eqns~\ref{eqn:tomo}-\ref{eqn:c1}). Unlike in photometric galaxy clustering studies, taking $\ell$-cuts can not be used to completely mitigate bias from poorly understood nonlinear scales. Second, the mixing of scales in the radial direction is not a consequence of the plane parallel approximation. This can be seen by taking the high-$\ell$ limit where the plane parallel power spectra are equivalent to the full non-Limber spherical power spectra~\cite{Gebhardt:2020imr}. Instead the mixing of scales is a direct consequence of the anisotropy of the RSD field. 
\par Let us suppose that we have a model of nonlinear RSD which is unbiased for all $k_\parallel < k_\parallel^{\rm max}$, for some $k_\parallel^{\rm max}$. This suggests that we should keep decreasing $\Delta z$ as long as $\widetilde K_{ij}(k_\parallel) \sim 0$ for all $k > k_\parallel^{\rm max}$. This ensures we capture information from all modes with $k_\parallel < k_\parallel^{\rm max}$, while removing sensitivity to all modes with $k_\parallel > k_\parallel^{\rm max}$.

\par With this choice of bin width, we still do not retain all the information available in $P(k, \mu)$. To see this note that each $k_\parallel$-mode yields independent information. In fact, many of the parameters of interest such as the galaxy bias, growth function and FoG velocity dispersion all impact $P(k,\mu)$ in different regions of $k_\parallel-k_\perp$ space. This is shown in Fig.~\ref{fig:derivatives}, where we plot the RSD power spectrum derivatives with respect to these parameters, normalized against the RSD power spectrum. Yet from Eqn.~\ref{eqn:c1} we see that the tomographic $C(\ell)$ are given as an integral over $k_\parallel$-modes weighted by the kernel $\widetilde K_{ij}(k_\parallel)$. These kernels are broad in $k_\parallel$ leading to a mixing of modes across a wide range of radial scales so that in projection over $k_\parallel$, detailed information about $P(k_\perp, k_\parallel)$ and hence cosmological parameters, is lost. This is also manifested in additional degeneracies between cosmological parameters in projection. We quantify these statement in Section~\ref{sec:fisher results}.

\subsection{Angular Radial-Harmonic Power Spectra}
\par As we have just argued, the primary problem with tomographic $C(\ell)$ is that the radial-mode efficiency kernels, $\widetilde K_{ij} (k_\parallel)$ are broad in $k_\parallel$. In this section, we will use the generalized power spectra formalism and find a set of weight functions, $\{ w(\eta, z) \}$ which ensure that $ \widetilde K(k_\parallel;\eta_a,\eta_b)$ are narrow in $k_\parallel$ and form a spanning set over $k_\parallel$. 
\par For the sake of argument, let us suppose the window $W(r[z])$ is flat in redshift. It follows from Eqns.~\ref{eqn:w1}, \ref{eqn:K1} and \ref{eq:fourier window} that to ensure the radial-mode efficiency kernels are narrow in $k_\parallel$, we should choose a set of weight functions, $\{ w(\eta, z) \}$, which are narrow in Fourier space. Meeting this criteria, the key insight of this paper is to note that the Fourier transform of a cosine is a delta function, inspiring the choice:
\begin{equation} \label{eq:weight}
    w(\eta, z) = \cos \left( \frac{2 \pi \eta r^{\rm ref }[z]}{\Delta r^{\rm ref}[z]}\right).
\end{equation}
Here, $r^{\rm ref}$ and $\Delta r^{\rm ref}$ are the comoving distance to the window and the width of the window assuming a fiducial reference cosmology. In a likelihood analysis this weight should be applied to the data only once, and when computing the theory vector, the weight remains invariant even as the background cosmology is changed while sampling cosmological parameter space. The pre-factors are chosen so that the weight function has a period of $\Delta r^{\rm ref} / \eta$. With this choice of weight function, we  refer to the resulting $C^{\eta_a \eta_b} (\ell)$ as the {\it radial-harmonic power spectra}.

\subsection{$k_\parallel$-mode Separation in the Radial-Harmonic Basis and the $P(k,\mu) - C^{\eta_a \eta_b} (\ell)$ Hybrid Estimator} \label{sec:hybrid}

\par
\begin{figure}[!hbt]
\includegraphics[width = \linewidth]{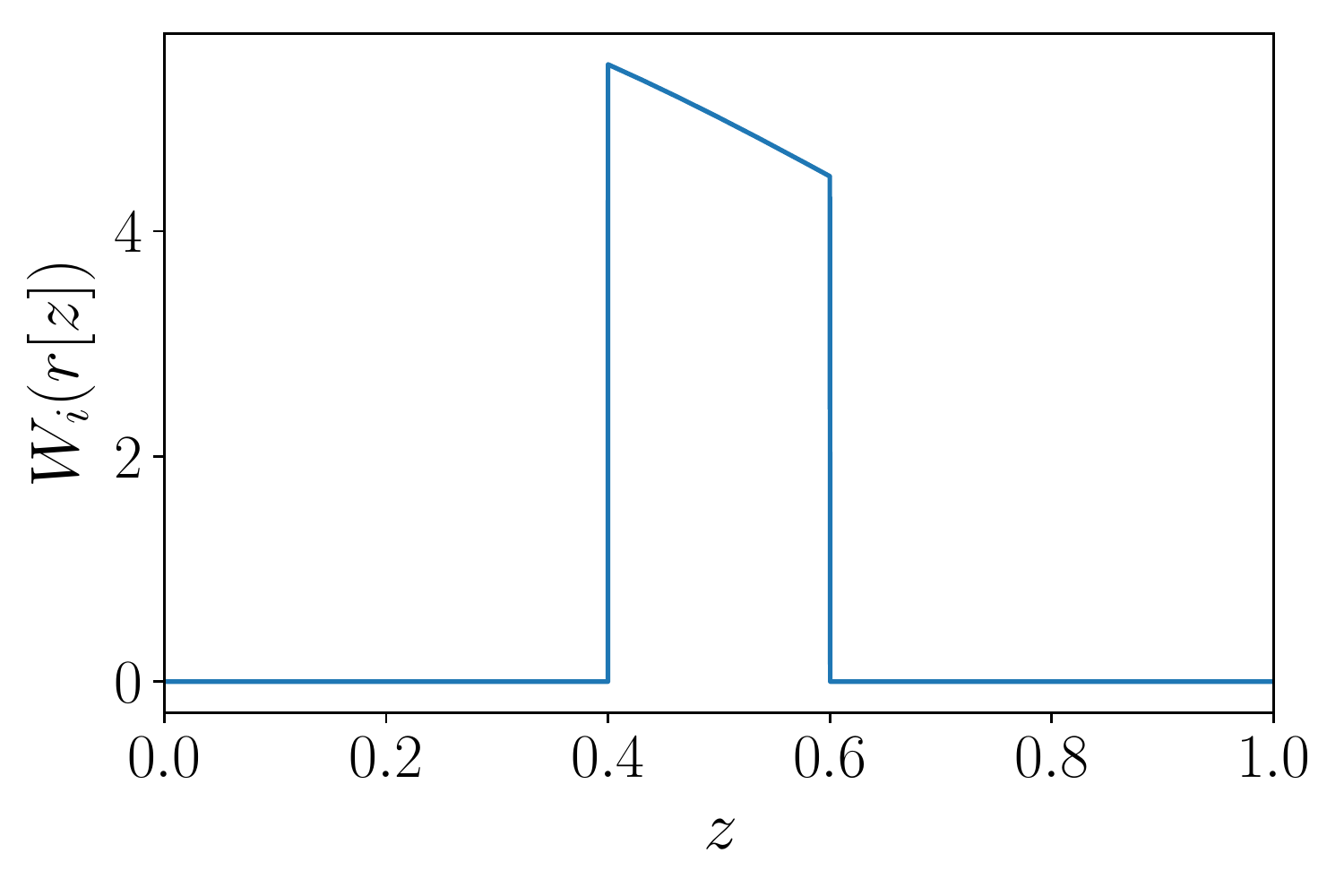}
\includegraphics[width = \linewidth]{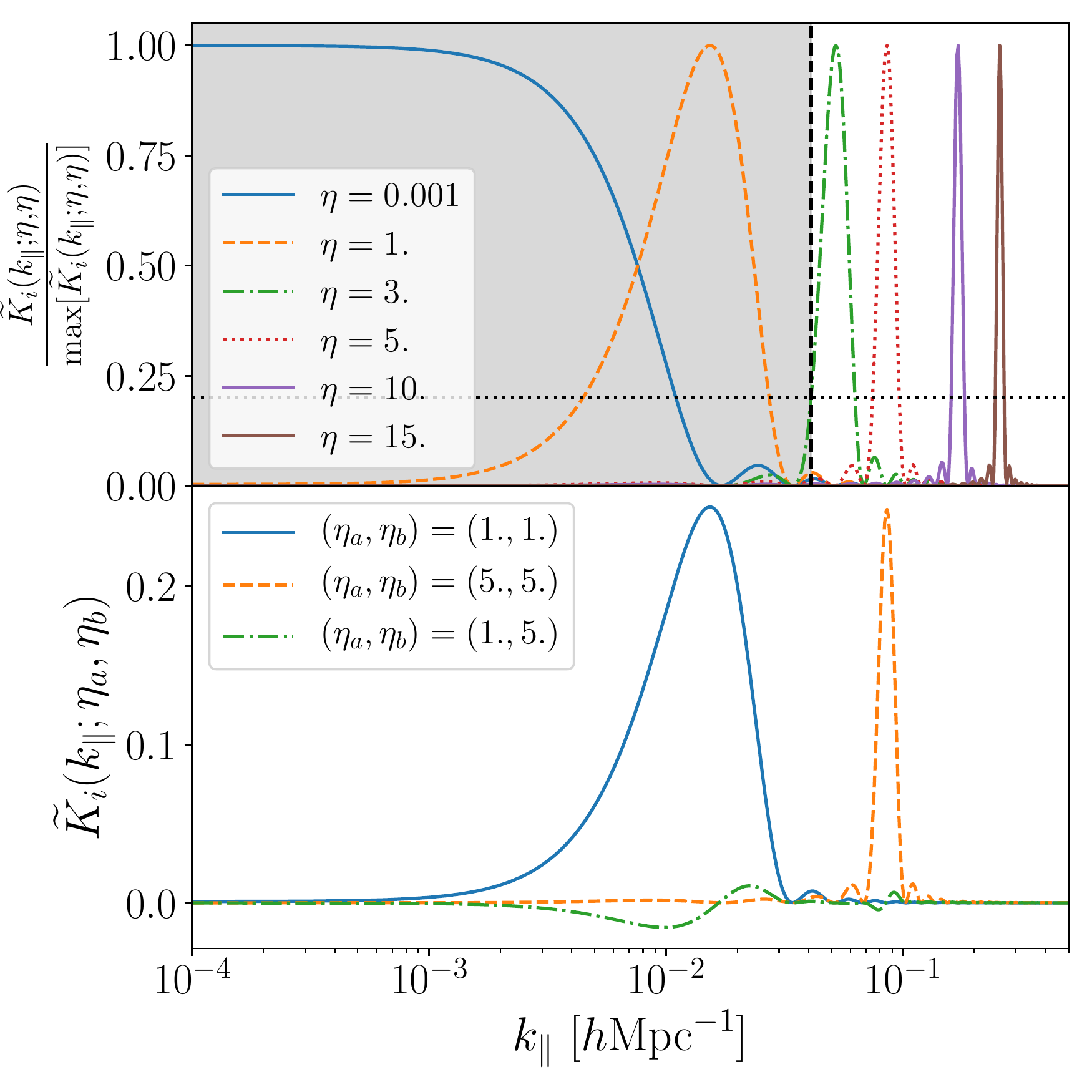}
\caption{{\bf Top:} The radial window function of a Stage IV-like survey. {\bf Middle:} The radial-mode efficiency kernel normalized against its maximal value. With the hybrid estimator we use $P(k_\parallel, k_\perp)$ to probe scales to the left of the black dashed line in grey and $C^{\eta_a \eta_b} (\ell)$ to probe scales to the right of the line. In this regime the kernels are narrow so there is a nearly one-to-one correspondence between $\eta$ and $k_\parallel$, implying that in contrast to the tomographic case (see Figure~\ref{fig:tomo_kernel}), radial $k_\parallel$ scales are unmixed in projection. {\bf Bottom:} The radial-mode efficiency kernels.  For $\eta_a \neq \eta_b$ the magnitude of the radial-mode efficiency kernel is much smaller than for the auto-correlations between $\eta$-modes. Hence for fixed $\ell$, $C^{\eta_a \eta_b} (\ell)$ is nearly diagonal.}
\label{fig:harmonic kernels}
\end{figure}

 In Figure~\ref{fig:harmonic kernels}, we plot the radial-mode efficiency kernels for a Stage IV-like survey centered on $z=0.5$.  The radial window, $W(r[z])$ is shown in the top panel. This is not flat in redshift to demonstrate that the weight function in Eqn.~\ref{eq:weight} works as intended in a more general setting. 

\par The radial-mode efficiency kernels normalized against their maximal values are shown in the middle row. As the value of $\eta$ is increased, the kernels become narrower and peak at higher values of $k_\parallel$. This means that in contrast to the tomographic power spectrum which mixes $k_\parallel$-scales, for $\eta \gtrsim 3$ the radially harmonic power spectrum, $C^{\eta \eta}(\ell)$, is only sensitive to $k_\parallel$-modes inside a narrow range and we probe different $k_\parallel$ scales for different values of $\eta$. Very roughly the scaling goes as
\begin{equation} \label{eqn:scaling}
k_\parallel \sim \frac{ 2 \pi \eta}{  \Delta r}.
\end{equation}
In fact, for $\eta \gtrsim 3$ ($k_\parallel \gtrsim 0.05 \ h {\rm Mpc}^{-1}$) there is a nearly a one-to-one correspondence between $k_\parallel$ and $\eta$. From Eqn.~\ref{eqn:c2a} there is also a one-to-one correspondence between $\ell$ and $k_\perp$. Thus nearly all the cosmological information that exists in the RSD power spectrum $P(k_\parallel, k_\perp)$ is retained after projection to  $C^{\eta_a \eta_b}(\ell)$ -- at least in the regime where $k_\parallel \gtrsim 0.05 \ h {\rm Mpc}^{-1}$. 
\par As $\eta \rightarrow 0$, we approach the tomographic case as the wavelengths of the radial weighting become much larger than the window. The fact that the kernels are broad and mix $k_\parallel$ scales for low-$\eta$ (low-$k_\parallel$) motivates the choice of a new $P(k,\mu) - C^{\eta_a \eta_b} (\ell)$ hybrid estimator. We shall refer to this as the {\it radial-harmonic hybrid estimator}. In the high-$k_\parallel$ (high-$\eta$) regime we will use the projected radial-harmonic power spectra $C^{\eta_a \eta_b}(\ell)$, enabling a cross-correlation with lensing at high-$k_\parallel$ to constrain nonlinear RSD. Meanwhile on large radial scales, we will use the RSD power spectrum $P(k_\parallel, k_\perp)$. 
\par We combine the two estimators as follows:
\begin{itemize}
\item Choose a minimum $\eta$-value, $\eta^{\rm min}$. This choice is not formally defined, but it should be as small a value as possible while ensuring that the radial-mode efficiency kernel, $\widetilde K(k_\parallel;\eta,\eta)$ is narrow.
\item Using the correspondence between $\eta$ and $k_\parallel$, define a scale $k_\parallel^{\rm max} (\eta = \eta^{\rm min})$ which is the largest value of $k_\parallel$ such that $\widetilde K(k_\parallel;\eta^{\rm min},\eta^{\rm min})$ is `small'. This ensures that $P(k_\parallel, k_\perp)$ and $C^{\eta_a \eta_b} (\ell)$ probe different $k_\parallel$ scales so that there is negligible covariance between the two. For the remainder of this paper we will take $k_\parallel^{\rm max} (\eta^{\rm min})$ as the maximum $k_\parallel$ such that $K(k_\parallel;\eta ^{\rm min},\eta^{\rm min} )$ is less than $20 \%$ of its maximal value over $k_\parallel$, corresponding to the horizontal dashed line in the middle row of Fig.~\ref{fig:harmonic kernels}.
\end{itemize}
\par Following this procedure, if we choose $\eta^{\rm min} = 3$, then $k_\parallel^{\rm max} (\eta^{\rm min}) = 0.04$ which is indicated by the dotted black line in the middle panel of Figure~\ref{fig:harmonic kernels}. The grey shaded region to the left of this line indicates the $k_\parallel$ modes which are probed with $P(k_\parallel, k_\perp)$ and the unshaded region denoted scales which are probed with $C^{\eta_a \eta_b} (\ell)$.
\par The bottom panel of Figure~\ref{fig:harmonic kernels} shows the radial-mode efficiency kernels for $(\eta_a, \eta_b) = (1.,1.), (5.,5.), (1., 5.)$.  We see that for $\eta_a \neq \eta_b$, the magnitude of the radial-mode efficiency kernel is much smaller than for the auto-correlations between $\eta$-modes so that for fixed $\ell$, $C^{\eta_a \eta_b} (\ell)$ is nearly diagonal.

\subsection{Comparison of Radial-harmonic and Tomographic Power Spectra}

\par We start by validating our tomographic code against {\tt Cosmosis}~\cite{Zuntz:2014csq}. For this test we assume no anisotropic RSD contributions, i.e we set $\widetilde{A}^2_{\mathrm{RSD}} = 1$ in Eqn.~\ref{eq:pkmu} and assume a top-hat window in the range $z \in [0.45, 0.55]$. All data points agree within $5 \%$ with average agreement within $1 \%$ for $\ell \in [100, 1000]$. {\tt Cosmosis} makes the Limber approximation~\cite{LoVerde:2008re} and assumes a temporally evolving power spectrum while we assume that the power spectrum is fixed inside the tomographic bin and assume the plane parallel approximation~\cite{Jalilvand:2019brk}. In both cases we use {\tt Camb}~\cite{Lewis:1999bs} to generate the linear power spectrum and {\tt Halofit}~\cite{Takahashi:2012em} to generate the nonlinear power spectrum, which we call from {\tt PyCamb} in our code.
\par
\begin{figure}[!hbt]
\includegraphics[width = \linewidth]{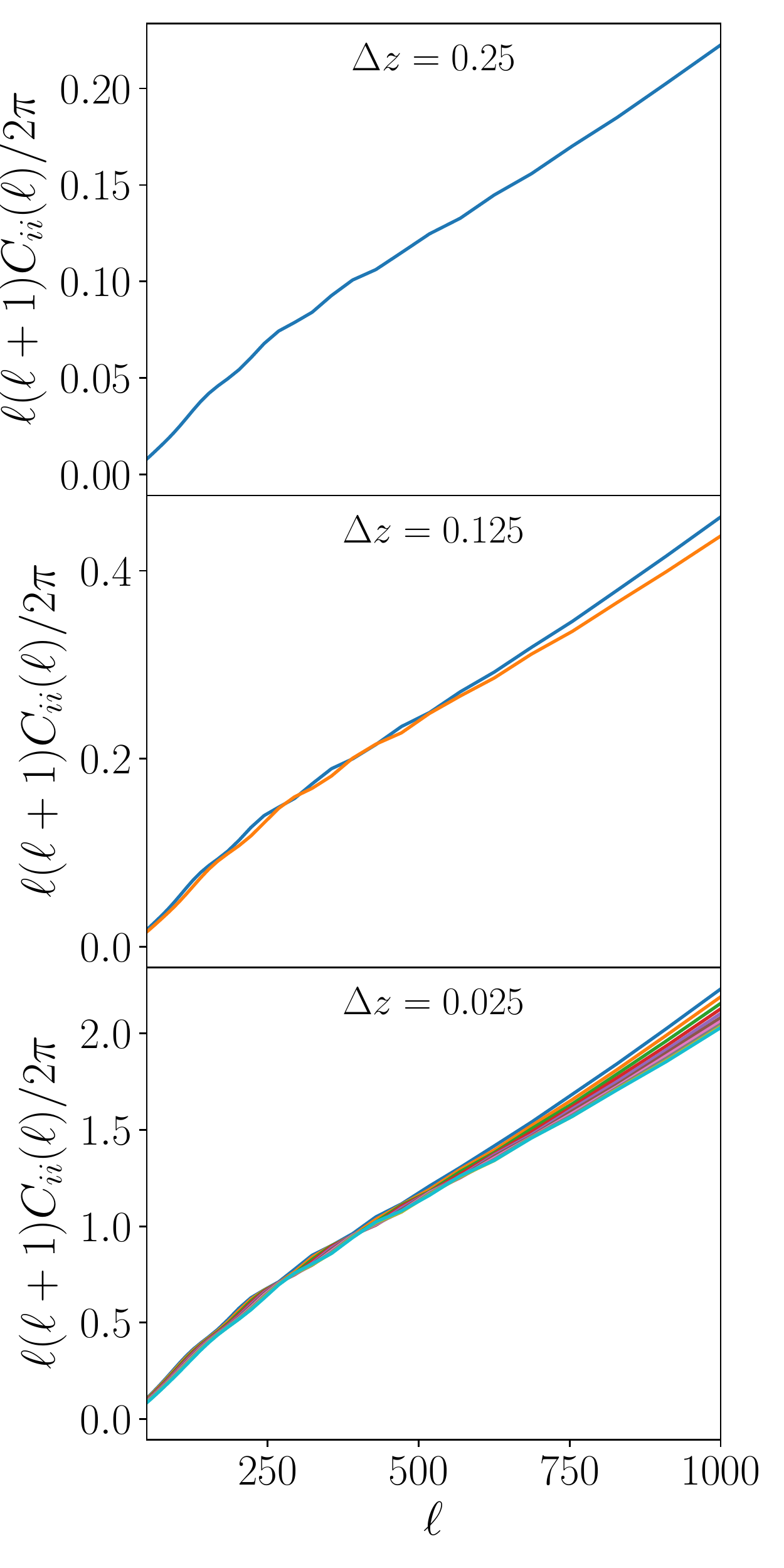}
\caption{Tomographic $C(\ell)$ in the redshift range $z \in [0.5, 0.75]$. Top, middle and bottom panels correspond to $1$, $2$ and $10$ tomographic subdivisions, corresponding to tomographic bin-width, $\Delta z$, of $0.25$, $0.125$ and $0.025$ respectively. The $C_{ii} (\ell)$ of narrow bins have more power as these functions probe high-$k_\parallel$ modes in addition to low-$k_\parallel$ modes (see Fig.~\ref{fig:tomo_kernel}). For a given tomographic bin width, the $C(\ell)$ of each tomographic bin probe approximately the same scales so that  $C_{ii} (\ell) \approx C_{jj} (\ell)$ for $i \neq j$. This implies that if $C_{ii} (\ell)$ is biased for some $\ell$-mode due to modelling uncertainty, then so is $C_{jj} (\ell)$ for $i \neq j$.}
\label{fig:cl_tomo}
\end{figure}

 In Fig.~\ref{fig:cl_tomo} we plot the tomographic $C_{ii} (\ell)$ that cover the top hat window over the range $z \in [0.5, 0.75]$ for $1$, $2$ and $10$ subdivisions corresponding to tomographic bin-widths, $\Delta z$, of $0.25$, $0.125$ and $0.025$ respectively. 
\par Comparing the three panels of Fig.~\ref{fig:cl_tomo}, we notice that as the tomographic bin-width is decreased the magnitude of the $C_{ii} (\ell)$ increases. This is precisely as expected because on comparison with Fig.~\ref{fig:tomo_kernel}, it is clear that narrow tomographic bins are sensitive to high-$k_\parallel$ modes as well as the low-$k_\parallel$ modes which are probed with broader tomographic bins. It is important to note that for narrow tomographic bins, even though we must use all tomographic bin pairs to maximize the signal-to-noise, they are all sensitive to the same poorly modelled nonlinear $k_\parallel$-modes, even for low-$\ell$.

\par
\begin{figure}[!hbt]
\includegraphics[width = \linewidth]{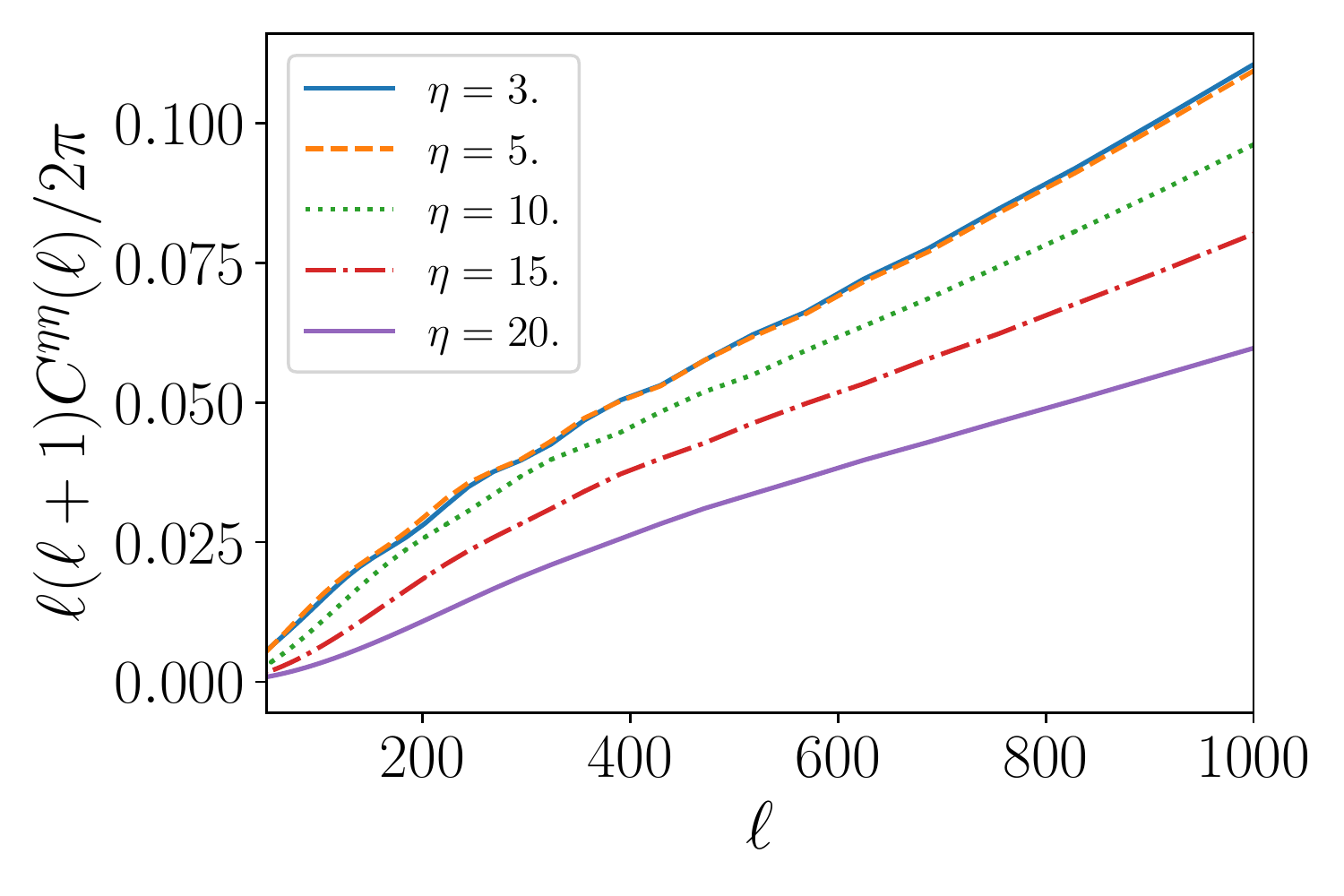}
\caption{Radial-harmonic power spectra inside the redshift range $z= [0.5, 0.75]$. Different values of $\eta$ probe different $k_\parallel$-modes resulting in differences between spectra in contrast to the tomographic case (see Fig.~\ref{fig:cl_tomo}).}
\label{fig:cl_harm}
\end{figure}

 In Fig.~\ref{fig:cl_harm} we show the radial-harmonic power for different values of $\eta$. Unlike in the tomographic case, the projected spectra probe different $k_\parallel$-scales at each $\eta$-value, resulting in differences in the power spectra. One can easily remove sensitivity to poorly modelled $k_\parallel$-modes by taking an $\eta$-cut to the data vector. For some fixed $\ell$-mode, this preserves useful information at low-$\eta$ while removing data points which are biased due to nonlinear RSD modelling uncertainty at high-$\eta$.

\par
\begin{figure}[!hbt] 
\includegraphics[width = \linewidth]{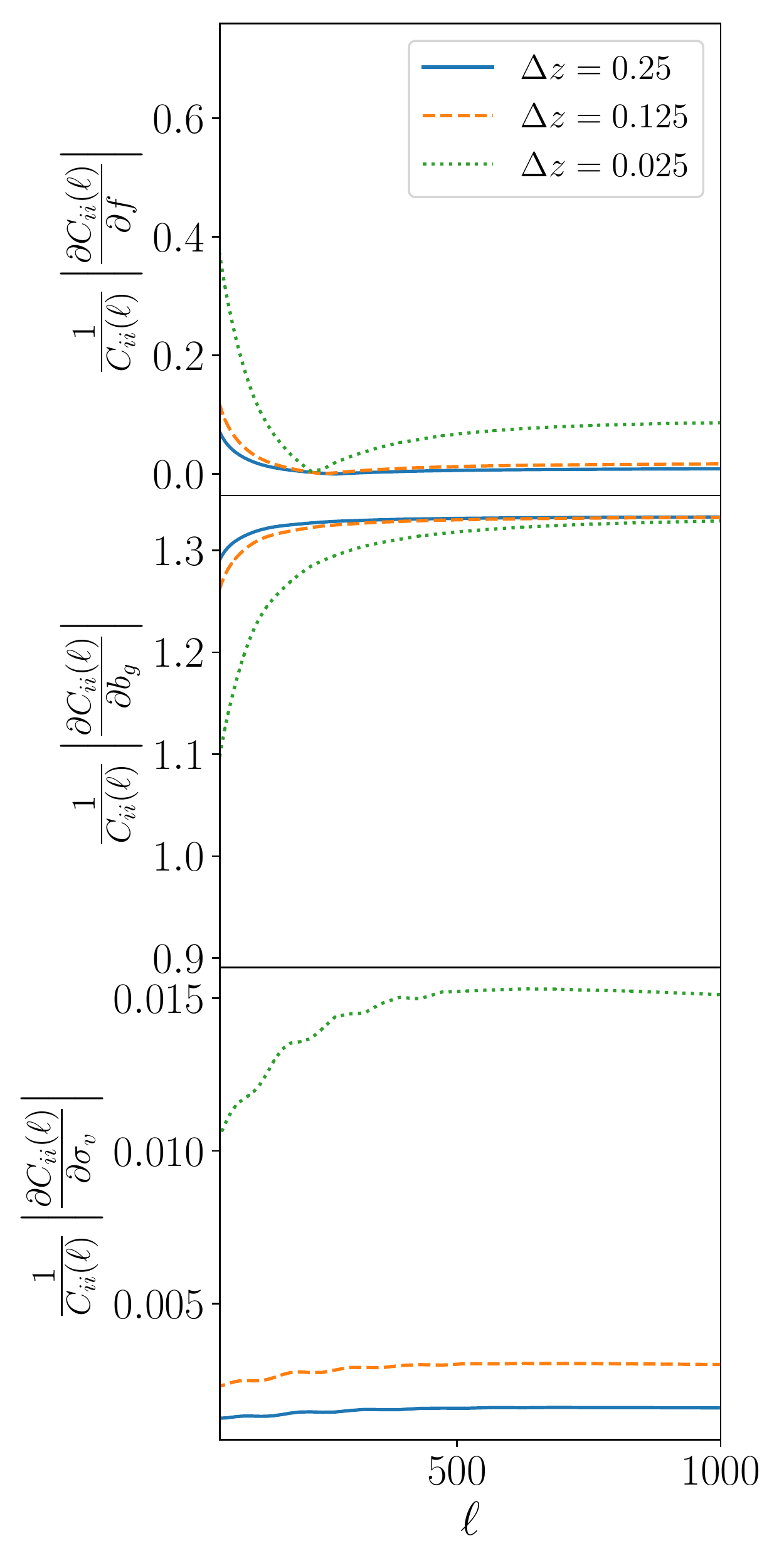}
\caption{The absolute value of the derivatives of the tomographic power spectra normalized against the power spectrum. This is a measure of how sensitive the tomographic power spectra are to different parameters (see Sec.~\ref{sec:fisher form}). The top row indicates that narrow tomographic bins are the most sensitive to $f$. However, from the bottom row, narrow tomographic bins are the most sensitive to $\sigma_v$, and more generally nonlinear RSD modelling uncertainties, so this binning strategy is the most prone to model bias.}
\label{fig:cl_tomo_deriv}
\end{figure}

 For parameters $f$, $b_g$ and $\sigma_v$, in Fig.~\ref{fig:cl_tomo_deriv} we plot the absolute value of the derivatives of the tomographic power spectra normalized against the power spectrum. This is a measure of how sensitive the tomographic power spectra are to these parameters (see Sec.~\ref{sec:fisher form} for more details). As before we consider tomographic bins widths, $\Delta z$, of $0.25$, $0.125$ and $0.025$. In each case we only consider the lowest redshift tomographic bin, since the differences between each bin's power spectra are small (see Fig.~\ref{fig:cl_tomo}).
\par The normalized derivative in the top panel of Fig.~\ref{fig:cl_tomo_deriv} is largest for narrow tomographic bins implying that the sensitivity to the high $k_\parallel$-modes yields additional constraining power on $f$. From the bottom plot in Fig.~\ref{fig:cl_tomo_deriv} we find that the power spectra of narrower bins are significantly more sensitive to, $\sigma_v$, as these power spectra probe high $k_\parallel$-modes (see Fig.~\ref{fig:derivatives}). More generally data points that are sensitive to $\sigma_v$ are those which are most prone to nonlinear RSD model bias. This plot shows that although narrow bins would yield the tightest constraints on $f$, this comes at the cost of model bias. 

\par
\begin{figure}[!hbt]
\includegraphics[width = \linewidth]{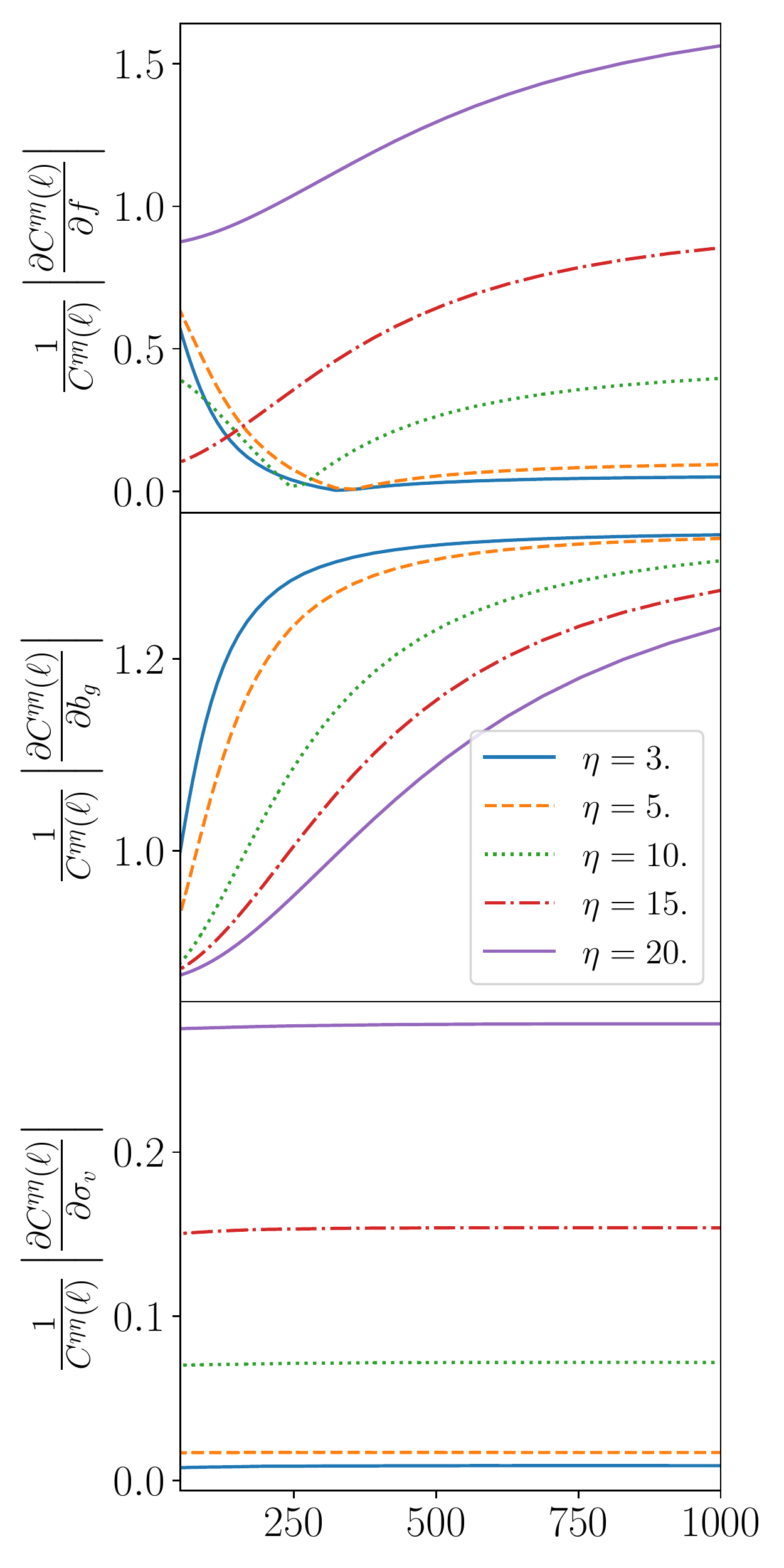}
\caption{Same as Fig.~\ref{fig:cl_tomo_deriv}, but for the radial-harmonic power spectra. From the top row all power spectra are sensitive to $f$. Crucially from the bottom row, only the high-$\eta$ power spectra are sensitive to $\sigma_v$ and more generally nonlinear RSD modelling uncertainties. These data points can be easily removed from the data vector, while preserving data points at low-$\eta$ which constrain the growth function, $f$. This is in contrast to the tomographic case, where we choose a tomographic bin width at the outset so that all data points are affected by modelling uncertainties.}
\label{fig:cl_harm_deriv}
\end{figure}

 Finally the normalized derivatives of the radial-harmonic power spectra are shown in Fig.~\ref{fig:cl_harm_deriv}. In this case we find that at low-$\ell$, the lower value $\eta$-modes are sensitive to $f$. Crucially these data points are extremely insensitive to $\sigma_v$ and nonlinear RSD scales in general, so the model bias is small. Meanwhile from the bottom panel of Fig.~\ref{fig:cl_harm_deriv} it is evident that we are most sensitive to $\sigma_v$, and nonlinear RSD in general, for large-$\eta$. These data points can easily be removed from the data vector to mitigate model bias, while constraining $f$ with the data points at low-$\eta$.

\section{Fisher Results} \label{sec:fisher results}
\subsection{Fisher Formalism} \label{sec:fisher form}
To compare the constraining power and model bias of the anisotropic power spectrum, tomography and the radial-harmonic hybrid estimator, we will perform a Fisher analysis.  For this analysis we will assume a Gaussian covariance neglecting both the Super-Sample Covariance (SSC)~\cite{Takada:2013wfa,Lacasa:2018hqp, Barreira:2017fjz} and Non-Gaussian (NG)~\cite{Barreira:2017fjz,Sato:2013mq} terms. 
\par Given a set of model parameters, $\{ \theta_\alpha\}$, if we assume the data vector follows a Gaussian likelihood and that it is linear in the model parameters, then a good estimate of the marginal error on $\theta_\alpha$, is
\begin{equation}    
    \sigma(\theta_\alpha) = \sqrt{(F^{-1})_{\alpha \alpha}},
\end{equation}
where $F$ is the Fisher matrix.
For tomographic $C(\ell)$ the Fisher matrix is~\cite{Blanchard:2019oqi}
\begin{equation} \label{eq:fish2}
\begin{aligned}
F^{\rm tomo}_{\alpha \beta}=\sum_{\ell=\ell_{\min }}^{\ell_{\max }} \sum_{i j, m n} \frac{\partial C_{i j}(\ell)}{\partial \theta_{\alpha}}\left[\Delta C^{-1}(\ell)\right]_{j m} \\ \times \frac{\partial C_{m n}(\ell)}{\partial \theta_{\beta}}\left[\Delta C^{-1}(\ell)\right]_{n i},
\end{aligned}
\end{equation}
where $\alpha$ and $\beta$ label the cosmological parameters, $i,j,m$ and $n$ label tomographic bins and,
\begin{equation} \label{eq:fish1}
    \Delta C_{i j}(\ell)=\sqrt{\frac{2}{(2 \ell+1) f_{\mathrm{sky}} \Delta \ell}}C_{i j}(\ell).
\end{equation}
Here $\Delta \ell$ is the multipole bandwidth and $f_{\mathrm{sky}}$ is the fraction of the sky covered by the survey. 
\par Let us suppose that we have two different nonlinear RSD models for the tomographic power spectra $C_{ij}(\ell)$ and we write the difference between the two models, $C^{m_1}_{ij} (\ell)$ and $C^{m_2}_{ij} (\ell)$, by defining
\begin{equation}
    \delta C_{ij} (\ell) = C^{m_1}_{ij} (\ell) - C^{m_2}_{ij} (\ell).
\end{equation}
Then for parameter $\theta_\alpha$, the expected bias between models, $b_\alpha$, is given by~\cite{Amara:2007as}
\begin{equation}
    b^{\rm tomo}_\alpha =\sum_\beta \left[ (F^{\rm tomo})^{-1}\right]_{\alpha \beta} B^{\rm tomo}_\beta,
\end{equation}
where
\begin{equation}
\begin{aligned}
    B^{\rm tomo}_\beta = \sum_{\ell=\ell_{\min }}^{\ell_{\max }} \sum_{i j, m n} \delta C_{ij} (\ell)  \left[\Delta C^{-1}(\ell)\right]_{j m} \\ \times \frac{\partial C_{m n}(\ell)}{\partial \theta_{\beta}}\left[\Delta C^{-1}(\ell)\right]_{n i}.
    \end{aligned}
\end{equation}

\par In terms of the RSD power spectrum the Fisher matrix is given by~\cite{Tegmark:1997rp}
\begin{equation}
\begin{aligned}
F^{3D}_{\alpha \beta} = \frac{V_s}{8 \pi ^2} \sum_{ k = k_{\rm min}}^{k_{\rm max}} \sum^{1}_{\mu = -1}  k^2   \frac{\partial P(k,\mu)}{\partial \theta_{\beta}}\  \frac{\partial P(k,\mu)}{\partial \theta_{\beta}} \\ \times {\rm Cov} \left[ P(k,\mu), P(k',\mu') \right], \\ 
\end{aligned}
\end{equation}
where $V_s$ is the survey volume and the RSD power spectrum covariance is 
\begin{equation}
\begin{aligned}
    {\rm Cov} \left[ P(k,\mu), P(k',\mu') \right] &= \frac{2}{N_k} \left[ P(k,\mu) + \frac{1}{\bar n_g}\right]^2 \\ &\times \delta (k-k')  \delta(\mu - \mu'),
\end{aligned}
\end{equation}
and the number of modes in the survey volume, $N_k$, is 
\begin{equation}
    N_k = \frac{k^2 \Delta k \Delta \mu }{4 \pi ^2}V_s.
\end{equation}
Assuming that we have two different nonlinear RSD models with power spectra $P^{m_1} (k,\mu)$ and $P^{m_2} (k,\mu)$, and we write
\begin{equation}
    \delta P(k,\mu) = P^{m_1} (k,\mu) - P^{m_2} (k,\mu),
\end{equation}
and 
\begin{equation}
\begin{aligned}
B^{3D}_{ \beta} = \frac{V_s}{8 \pi ^2} \sum_{ k = k_{\rm min}}^{k_{\rm max}} \sum^{1}_{\mu = -1}  k^2   \delta P(k,\mu) \  \frac{\partial P(k,\mu)}{\partial \theta_{\beta}} \\ \times {\rm Cov} \left[ P(k,\mu), P(k',\mu') \right], \\ 
\end{aligned}
\end{equation}
the bias between models, for cosmological parameter $\alpha$, is
\begin{equation}
 b^{3D}_\alpha =\sum_\beta \left[ (F^{3D})^{-1}\right]_{\alpha \beta} B^{3D}_\beta.
 \end{equation}
This result can be derived by following the derivation given in the Appendix of~\cite{Taylor:2006aw}, which computes the linear bias for any estimator with a Gaussian likelihood.
 \par For the angular radial-harmonic power spectra, $C^{\eta_a \eta_b} (\ell)$, we define  $F_{\alpha \beta}^{\rm harm}$, $B_\beta^{\rm harm}$ and $b_\beta^{\rm harm}$ exactly as in the tomographic case, but now $i$, $j$, $m$ and $n$ label the $\eta$-modes rather than the tomographic bins. 
 \par In the hybrid case we cut scales as described in Sec.~\ref{sec:hybrid} and compute $F^{\rm harm}_{\alpha \beta}$, $B^{\rm harm}_\beta$, $F^{\rm 3D}_{\alpha \beta}$ and $B^{\rm 3D}_\beta$ as described above using the new scale cuts. Using the fact that the two estimators used in the hybrid approach probe different scales, the hybrid Fisher matrix is
 \begin{equation}
     F^{\rm hybrid}_{\alpha \beta} = F^{\rm 3D}_{\alpha \beta} + F^{\rm harm}_{\alpha \beta},
 \end{equation}
 and the bias
 \begin{equation}
      b^{\rm hybrid}_\alpha = \sum_\beta \left[ (F^{\rm hybrid})^{-1}\right]_{\alpha \beta} B^{\rm hybrid}_\beta,
 \end{equation}
 where 
 \begin{equation}
     B^{\rm hybrid}_\beta = B^{\rm 3D}_\beta + B^{\rm harm}_\beta.
 \end{equation}
 
     \subsection{Fisher Results $z \in [0.5, 0.75]$}

 We use the Fisher formalism described in the previous section to compute the errors on $f$, $\sigma_v$ and $b_g$ for different $k_\parallel$-cuts, while simultaneously computing the linear model bias between the Gaussian and Lorentzian FoGs. The primary aim of this analysis is to determine how tightly the three estimators can constrain cosmology, and in particular the growth function, while keeping the nonlinear RSD model bias in check. Since the Gaussian and Lorentzian FoG models disagree at high-$k_\parallel$ (see Eqns.~\ref{eqn:gauss FoG} and~\ref{eqn:lorentz fog}), we use the linear bias between the two models as a proxy for RSD nonlinear model uncertainty.

\begin{figure*}[!hbt]
\includegraphics[width = 8cm]{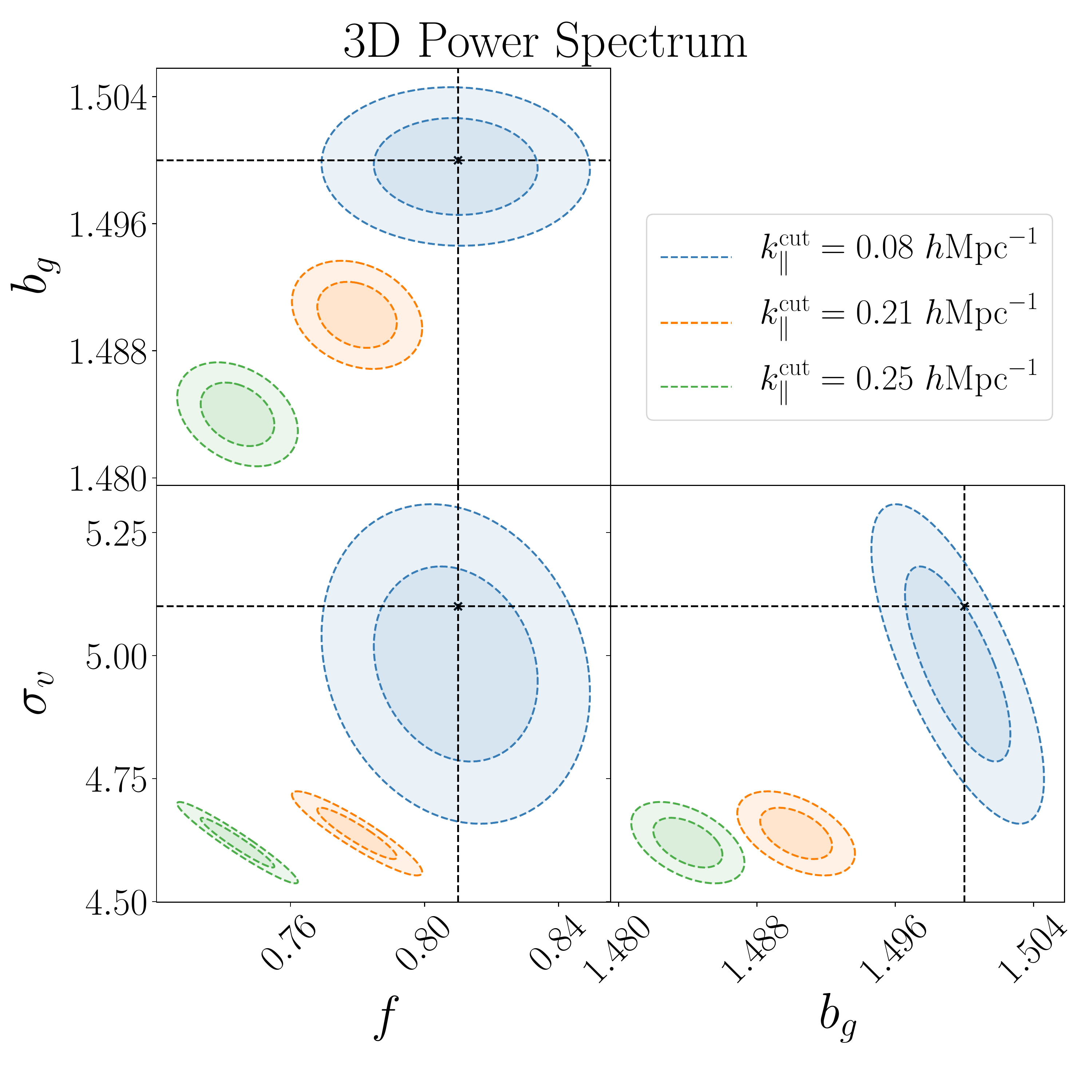}
\includegraphics[width =  8cm]{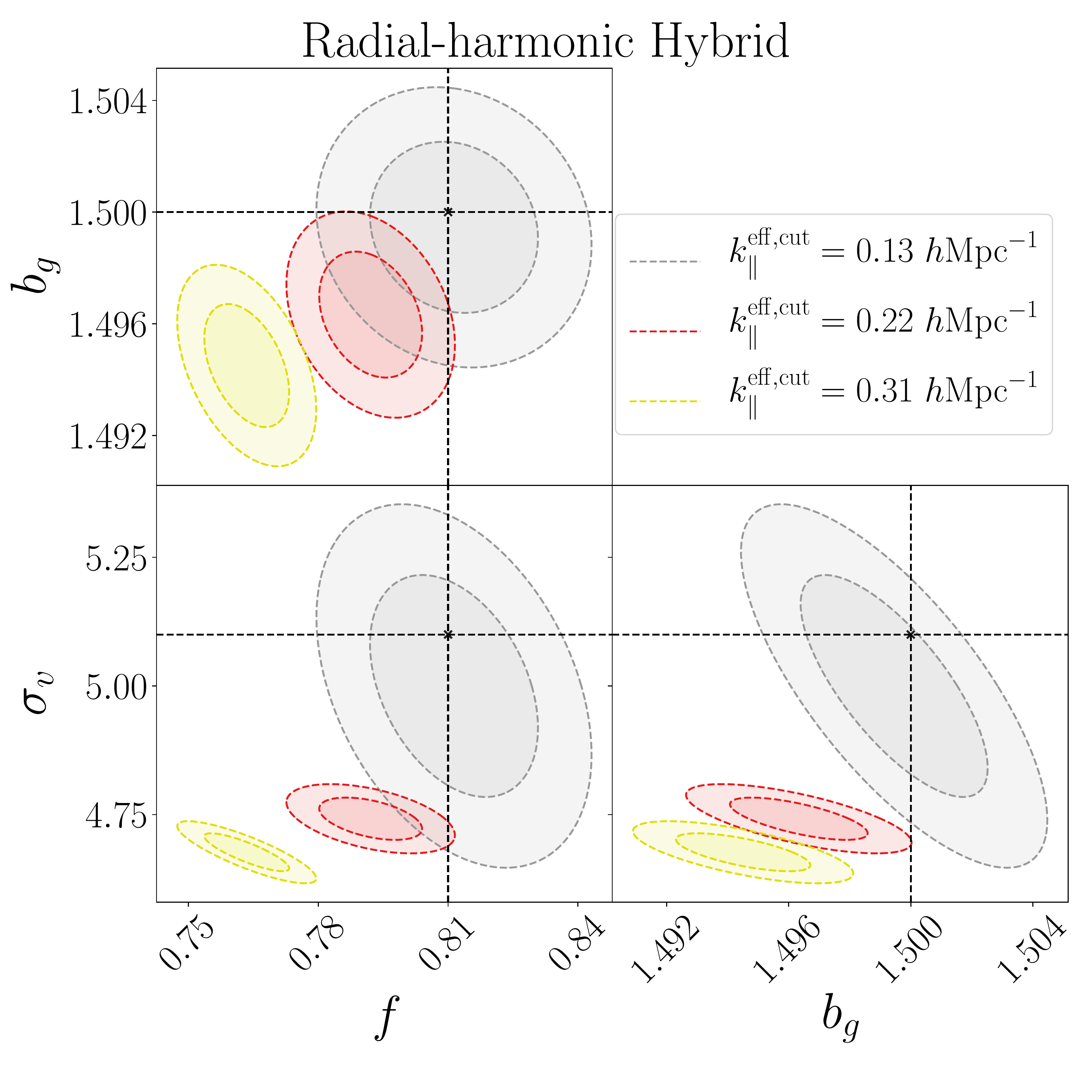}
\includegraphics[width = 8cm]{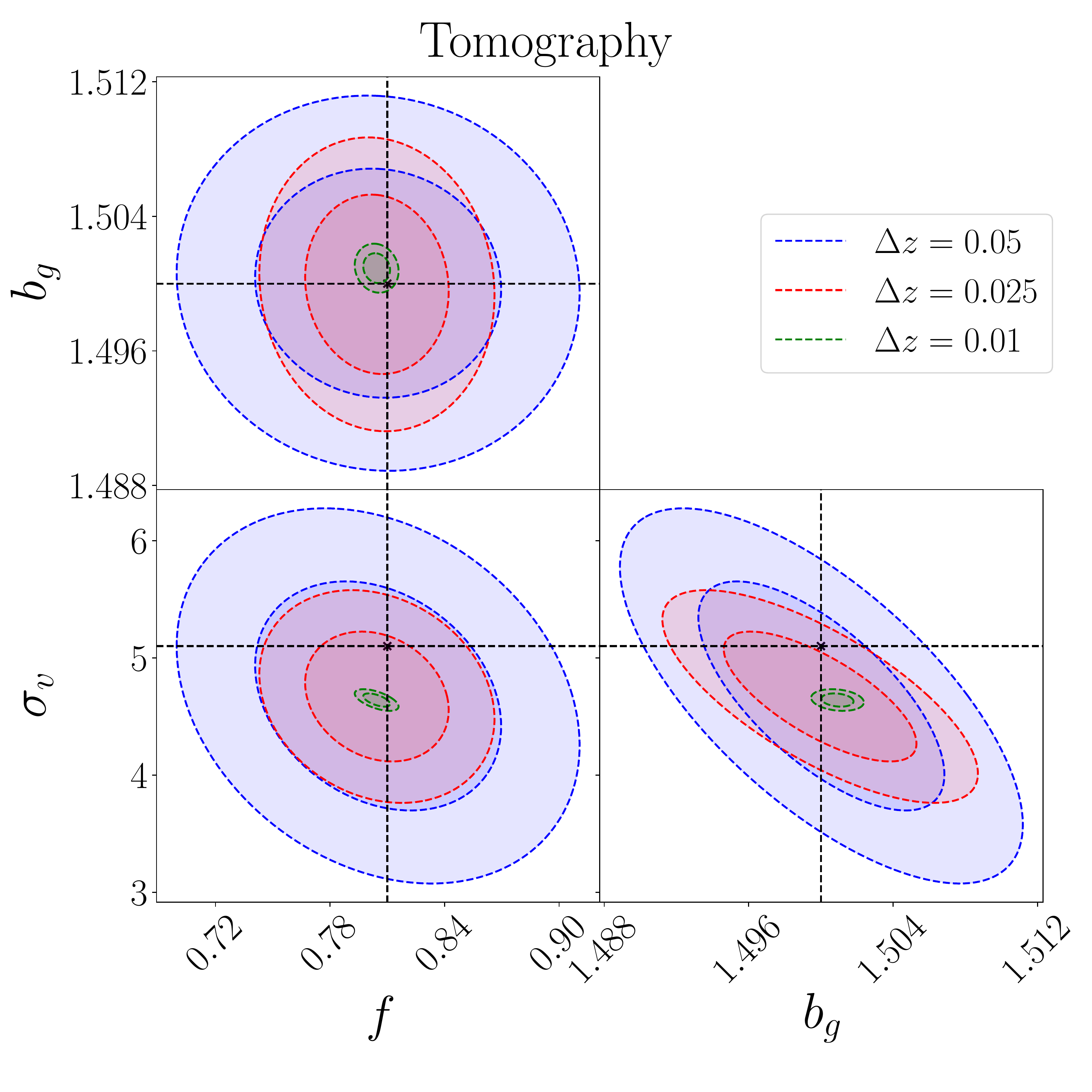}
\caption{The Fisher $68 \%$ and $95 \%$ confidence regions for tomography, the anisotropic power spectrum and the radial-harmonic hybrid estimator. {\bf The axes do no match for the three cases. For a direct comparison see Fig.~\ref{fig:fish compare}}. The contours are shifted by the linear model bias between the Lorentzian and Gaussian FoG simulating the impact of an incorrect nonlinear RSD model. The model bias increases with more optimistic scale cuts and as the precision of the constraints improves. This is particularly true for $\sigma_v$ which is the most sensitive to nonlinear scales.}
\label{fig:fish converge}
\end{figure*}

 \begin{figure*}[!hbt]
 \includegraphics[width = 5.5cm ]{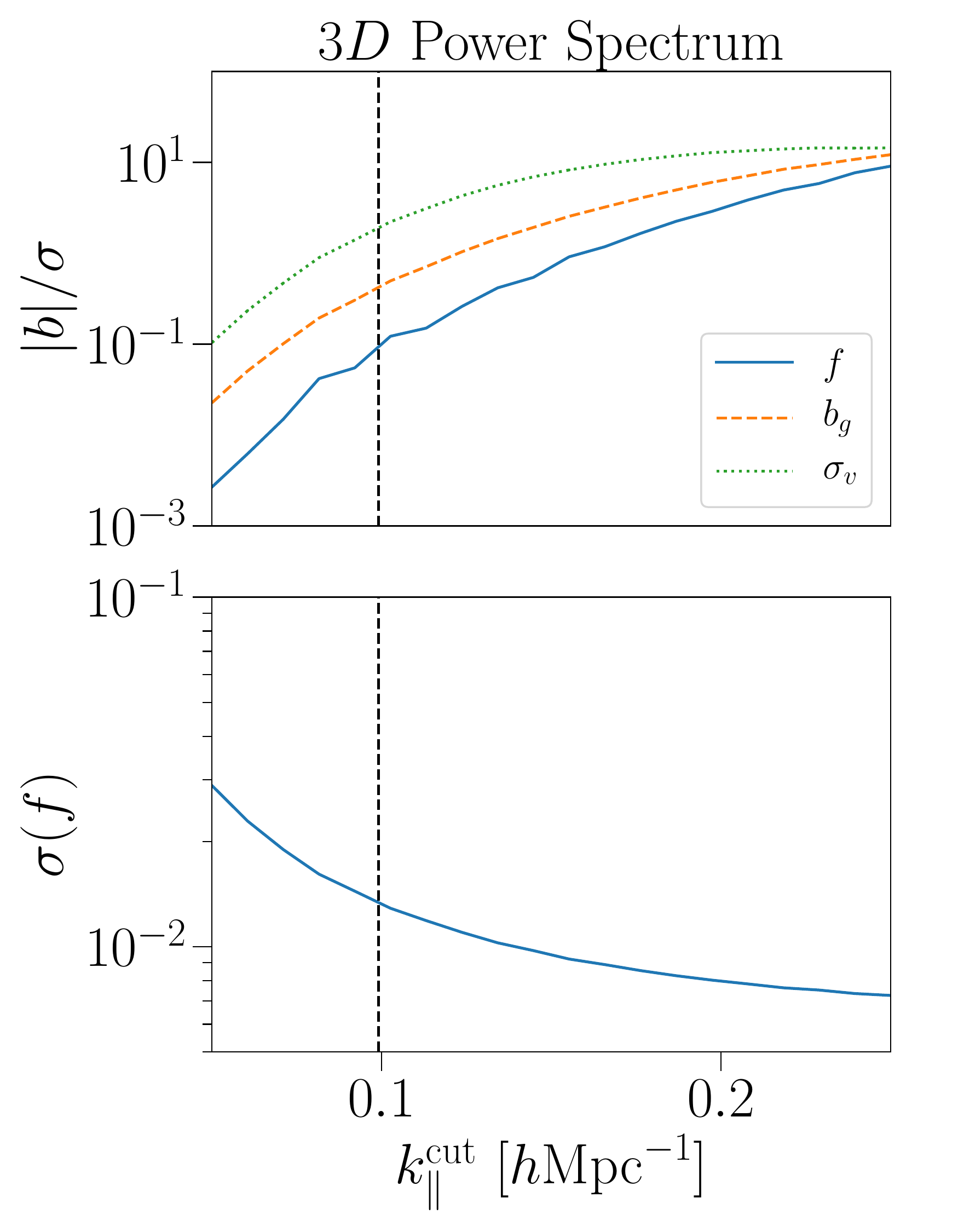}
 \includegraphics[width = 5.5cm ]{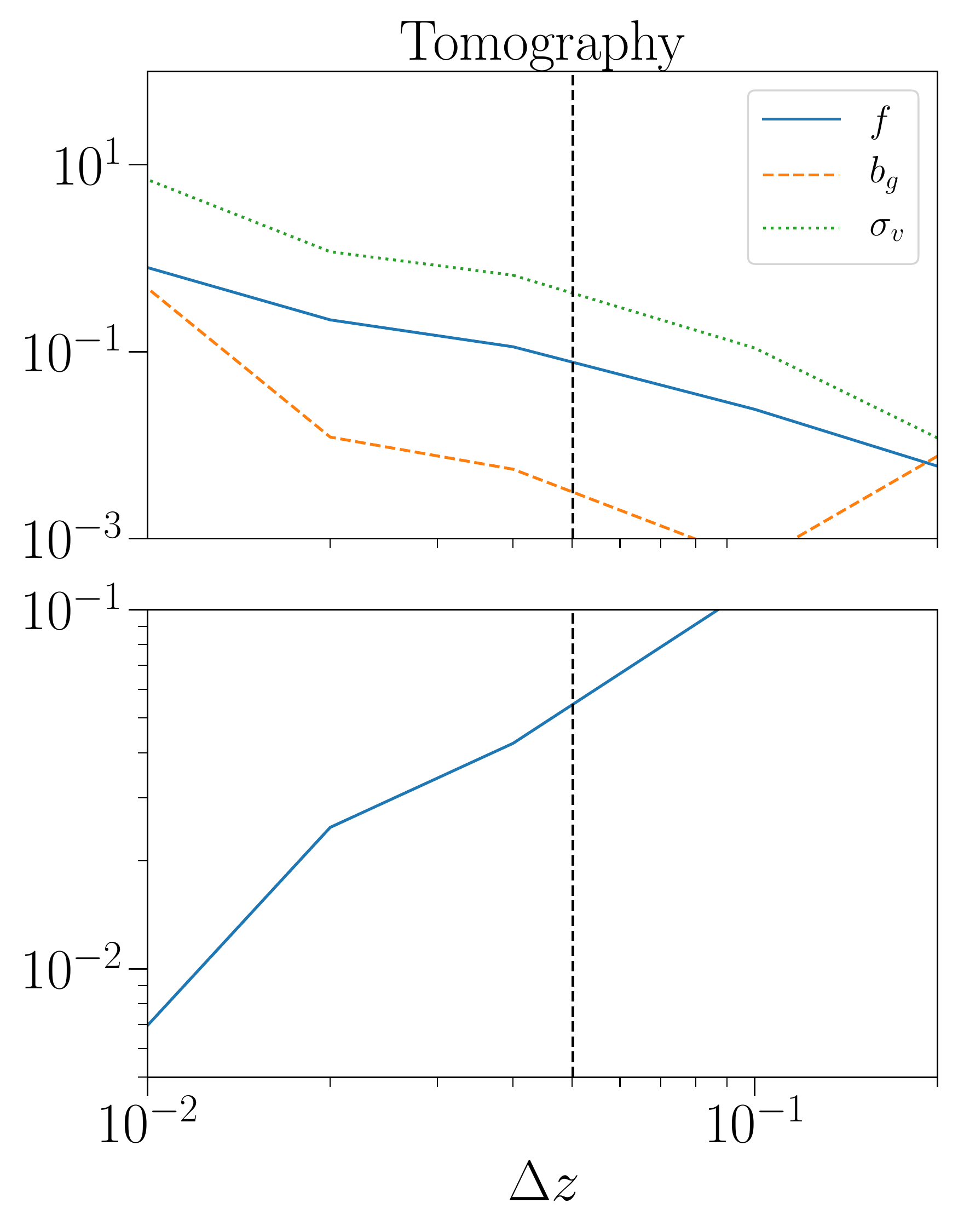}
 \includegraphics[width = 5.5cm ]{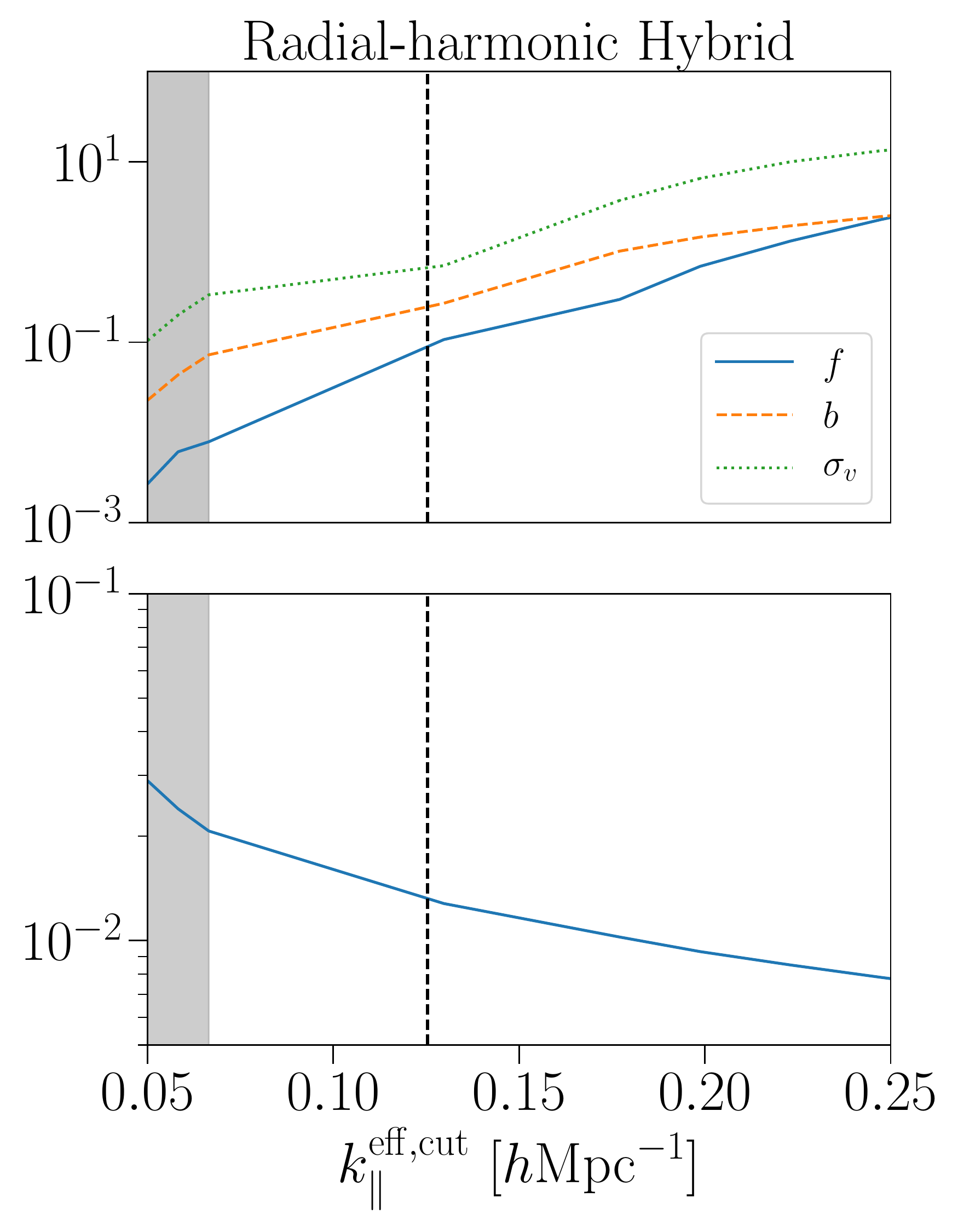}
\caption{{\bf Top:} The linear bias normalized against the Fisher error, $|b|/ \sigma$, as a function of cut scale. {\bf Bottom:} The error on the growth function, $\sigma (f)$. The dashed line indicates the maximum cut scale where the growth function constraints are unbiased in the sense that normalized bias is less than $10 \%$, i.e. $|b| / \sigma < 0.1$. This delineates our fiducial scale cuts. For this choice $\sigma(f) = 0.013$ for the anisotropic power spectrum, $\sigma(f) = 0.013$ for the radial-harmonic hybrid estimator and $\sigma(f) = 0.055$ for tomography. This amounts to a factor of $4$ degradation in constraining power when using tomography compared to the other two estimators. The full Fisher constraints are shown in Fig.~\ref{fig:fish compare}.}
\label{fig:results1}
\end{figure*}

 \begin{figure}[!hbt]
\includegraphics[width = \linewidth]{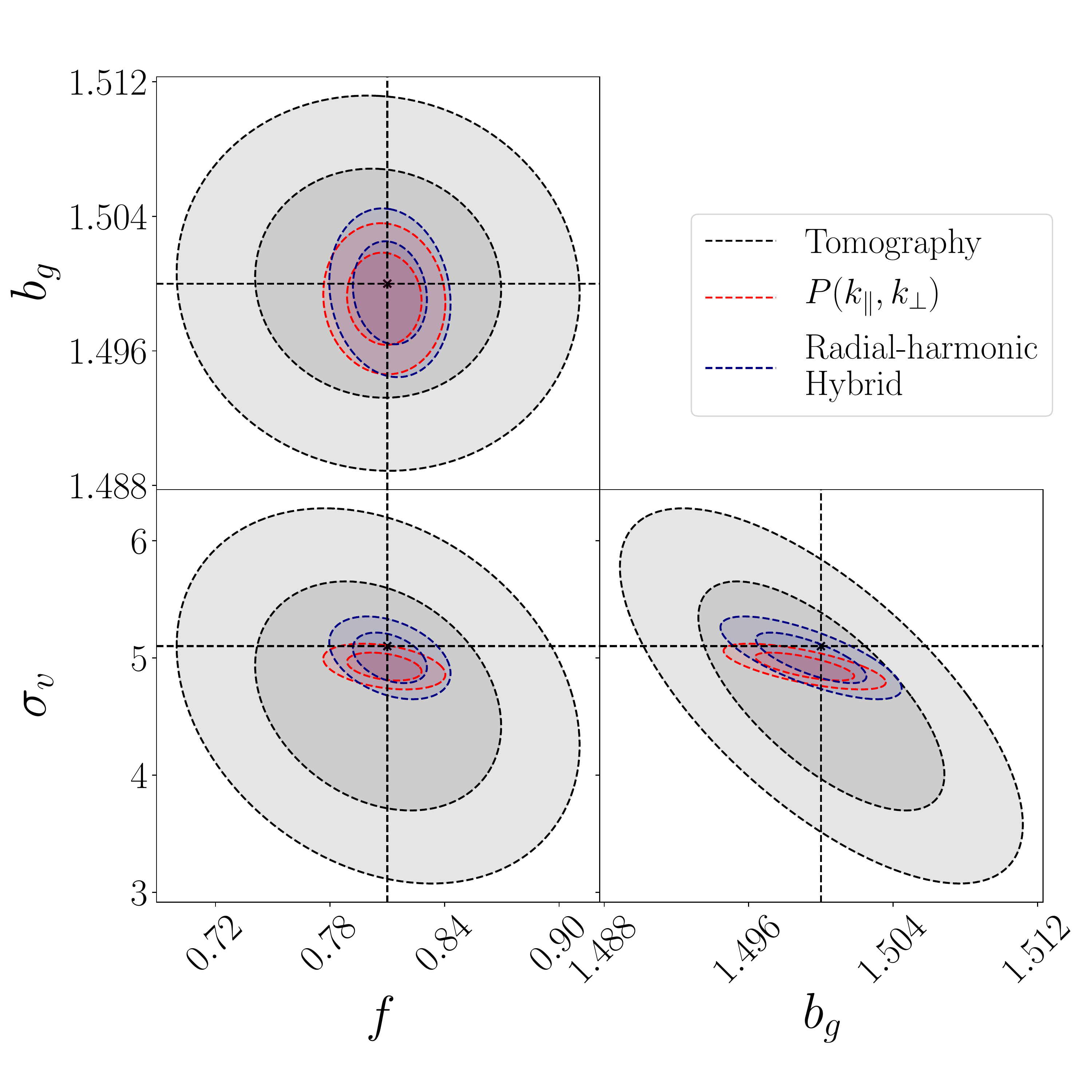}
\caption{Same as Fig.~\ref{fig:fish converge}, but now all three estimators are shown together to compare errors. We take the fiducial choice of scale cut, which maximizes precision while ensuring the bias on $f$ is closest to $10 \%$ (see Fig.~\ref{fig:fish compare} for more details). The radial-harmonic and anisotropic power spectrum constraints are nearly identical, while the tomographic errors are much larger due to the loss of information from mixing $k_\parallel$-scales in projection.}
\label{fig:fish compare}
\end{figure}

\par For this analysis we assume that $f_{\rm sky} = 0.3$ and that the radial window contains $7.5$ million galaxies uniformly distributed in the range $z \in [0.5, 0.75]$. This is chosen to match the redshift range of the highest-$z$ window in~\cite{Beutler:2016arn}, so that we can use the measured $\sigma_v$ from this study. For comparison the DESI~\cite{Aghamousa:2016zmz} catalogue will cover approximately a third of the sky and comprise of roughly $20$ million objects in the redshift range $z\in [0.5,1.5]$.\footnote{An additional bright galaxy sample BGS will comprise of an extra 10 million objects with a median redshift of $z \sim 0.2$.} We cut all perpendicular scales with $k_\perp > 0.25$, corresponding to $\ell > 350$ in the projected case. We take $\eta^{\rm min} = 5$, ensuring the radial-mode efficiency kernels are narrow.
\par The Fisher results as a function of the radial-mode scale cut are displayed in Fig.~\ref{fig:fish converge}. In the anisotropic power spectrum case, the scale cut is defined by a simple cut in $k_\parallel$, while for the radial-harmonic power spectrum, the cut corresponds to an $\eta$-cut which is converted to an effective cut-scale\footnote{For fixed $\eta$, we define the cut-scale, $k_\parallel^{\rm eff, cut}$, as the maximum $k_\parallel$ such that the radial-mode efficiency kernel is within $50 \%$ of it's maximal value. Since the kernels are narrow in $k_\parallel$-space (see Fig.~\ref{fig:harmonic kernels}) the precise way the cut is defined will have a negligible effect.}, $k^{\rm  eff, cut}_\parallel$. Meanwhile the scale for tomography is given by the bin width. Smaller bin widths correspond to more optimistic cuts in $k_\parallel$. 
\par From Fig.~\ref{fig:fish converge} we see that as expected, for each of the three estimators, the model bias increases with more optimistic scale cuts as the precision of the constraints improve. This is particularly true for $\sigma_v$ which is the most sensitive to nonlinear scales. 
\par A comparison of the three estimators is displayed in Fig.~\ref{fig:results1}. Each column corresponds to a different estimator. The grey shaded region in the hybrid case corresponds to the scales probed using the anisotropic power spectrum while the white regions indicate scale probed by the radial-harmonic power spectrum as in Fig.~\ref{fig:harmonic kernels}. In the first row of Fig.~\ref{fig:results1} we display the linear model bias normalized against the Fisher error, $|b|/ \sigma$. The error on the growth function after marginalising over $\sigma_v$ and $b_g$ is shown in the bottom row. While $|b|$ and $\sigma$ should  respectively increase and decrease monotonically with more optimistic scale cuts, $|b|/ \sigma$ is not necessarily strictly monotonic and it is not in the tomographic case.
\par We say that a parameter is {\it unbiased} if the normalized bias, $|b| / \sigma <0.1$, i.e. the bias is less than $10 \%$. This is similar to the criterion used in~\cite{massey2013origins} and corresponds to approximately a $95 \%$ overlap in the marginalized posterior probability distribution function~\cite{massey2013origins}. The scale cut at which the growth function, $f$, becomes biased is delineated by dashed line in Fig.~\ref{fig:results1}.
\par Since we are primarily concerned with the growth function, the parameter which distinguishes between gravitational models~\cite{Huterer:2013xky}, we take the point at which $f$ becomes biased as our fiducial choice of scale cut. In the anisotropic power spectrum case this corresponds to $k_\parallel^{\rm cut} \sim 0.1 \ h  {\rm Mpc}^{-1}$, in the tomographic case this corresponds to a tomographic bin width of $\Delta z = 0.05$ or $5$ tomographic bins covering the range $z \in [0.5, 0.75]$, and in the radial-harmonic case this corresponds to an effective cut at $k_\parallel^{\rm eff, cut} \sim 0.13 \ h  {\rm Mpc}^{-1}$. We do not expect the $k^{\rm cut}_\parallel$ in the anisotropic power spectrum case and $k^{\rm eff, cut}_\parallel$ in the radial-harmonic hybrid case to be identical. This is because the $k_\parallel$-$\eta$ correspondence is not absolute since the radial-mode, although narrow in $k_\parallel$-space by construction, still has some width. We have also been intentionally conservative when transitioning from $P(k_\parallel, k_\perp)$ to $C^{\eta_a \eta_b} (\ell)$ to ensure $k_\parallel$-modes are not double counted. This results in a small loss of information (see~\ref{sec:hybrid} for more details). 
\par In the second row of Fig.~\ref{fig:results1} we plot the error on the growth function, $\sigma(f)$, as a function of the cut scale. For our fiducial choice of scale (i.e $|b| \ \sigma <0.1$), $\sigma(f) = 0.013$ for the anisotropic power spectrum, $\sigma(f) = 0.013$ for the radial-harmonic hybrid estimator and $\sigma(f) = 0.055$ for tomography. This amounts to a factor of four degradation in constraining power when using tomography compared to the anisotropic power spectrum, caused by mixing of $k_\parallel$-scales in projection. Meanwhile the radial-harmonic power spectrum constraints are almost identical the anisotropic case demonstrating the harmonic weight works as intended. 
\par We plot the Fisher contours and bias for the three estimators in Fig.~\ref{fig:fish compare} assuming the fiducial scale cuts. The dotted line indicates the input cosmology, while the contours are shifted by the linear model bias between the Lorentzian and Gaussian FoG simulating the impact of using an incorrect nonlinear RSD model. Tomography is by far the least constraining for all parameters. The anisotropic power spectrum yield slightly tighter constrain on $\sigma_v$, but the constraints are otherwise indistinguishable.

\section{Fisher Results for $z \in [1,1.25]$}

 \begin{figure*}[!hbt]
\includegraphics[width = 5.5cm ]{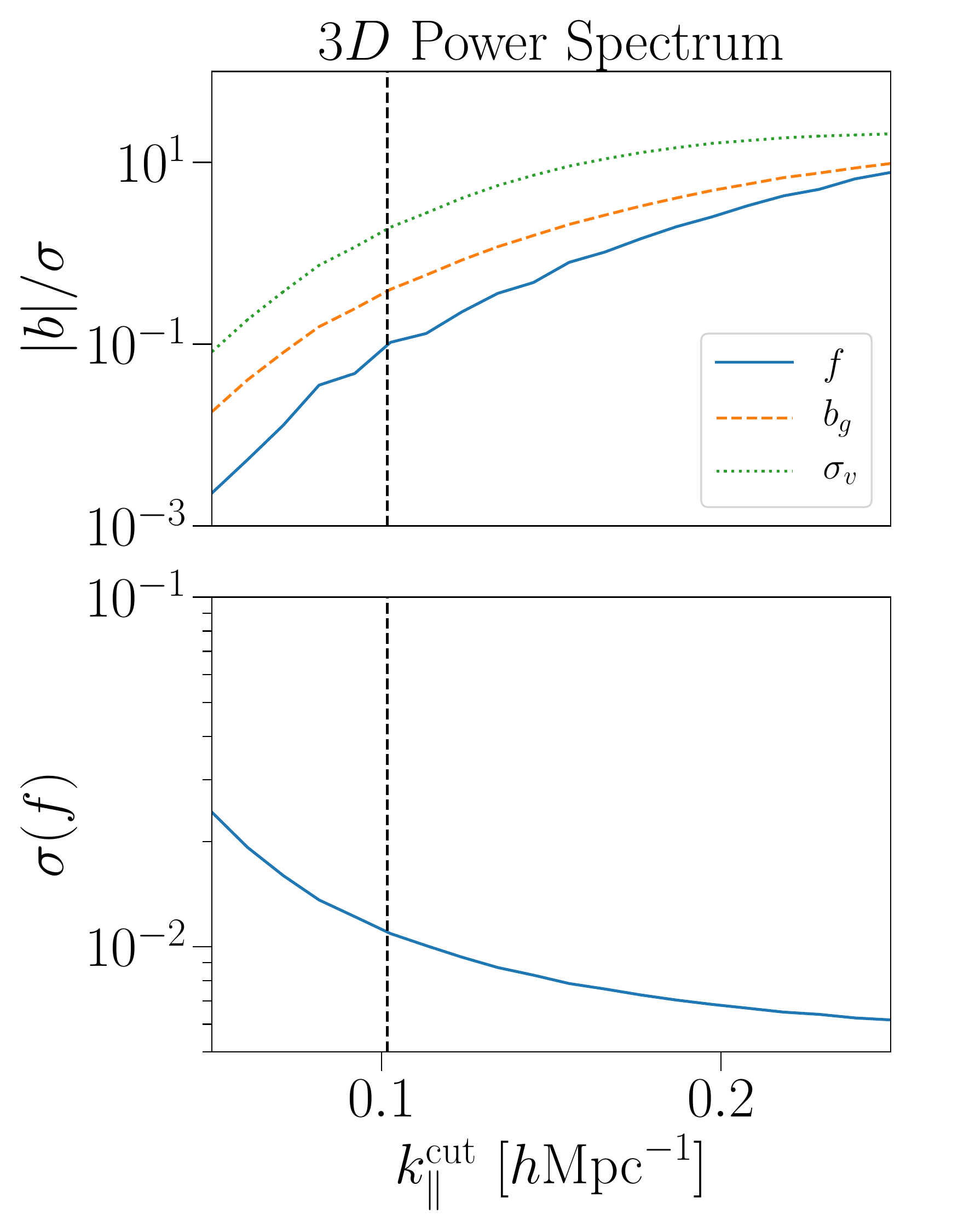}
\includegraphics[width = 5.5cm ]{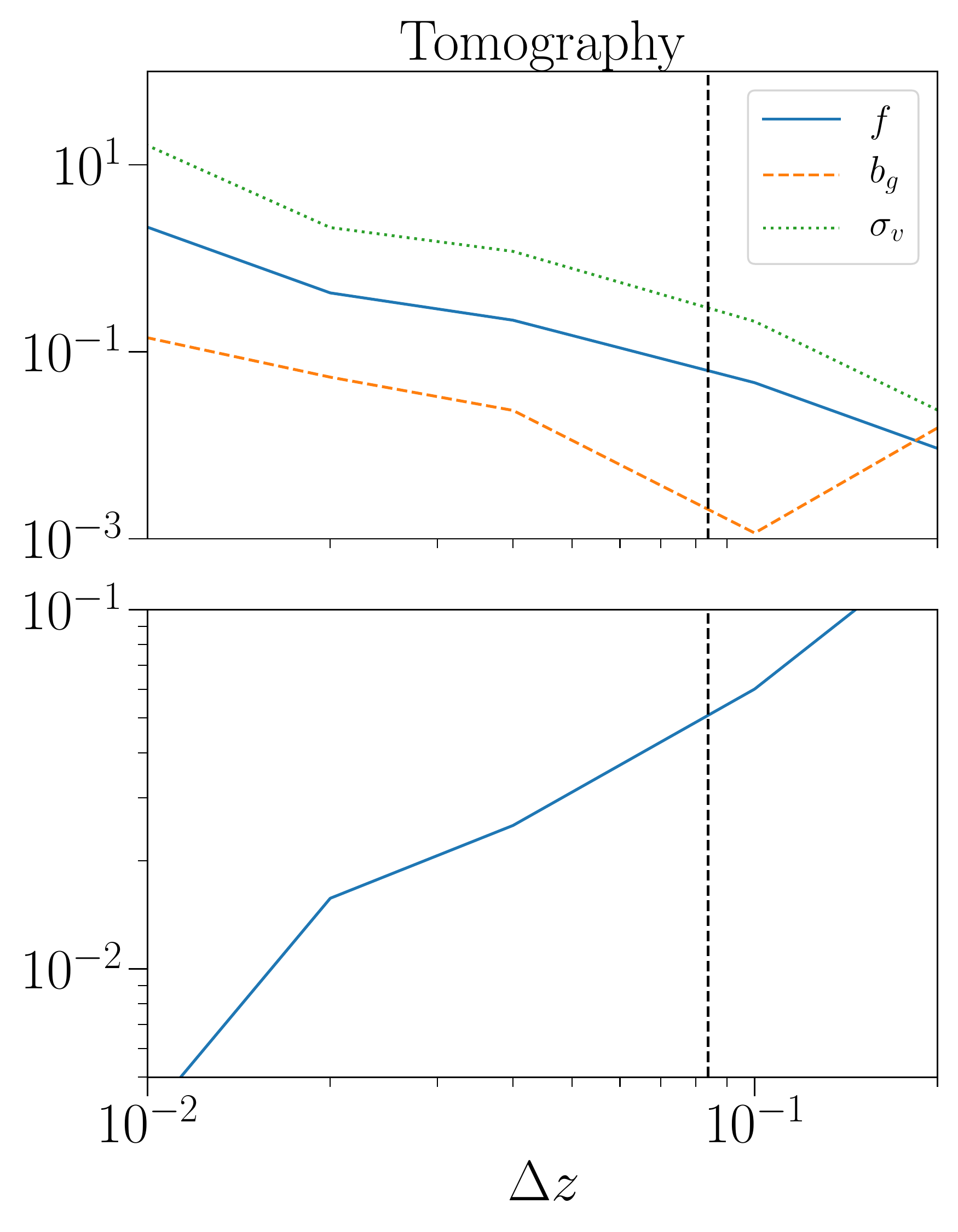}
\includegraphics[width = 5.5cm ]{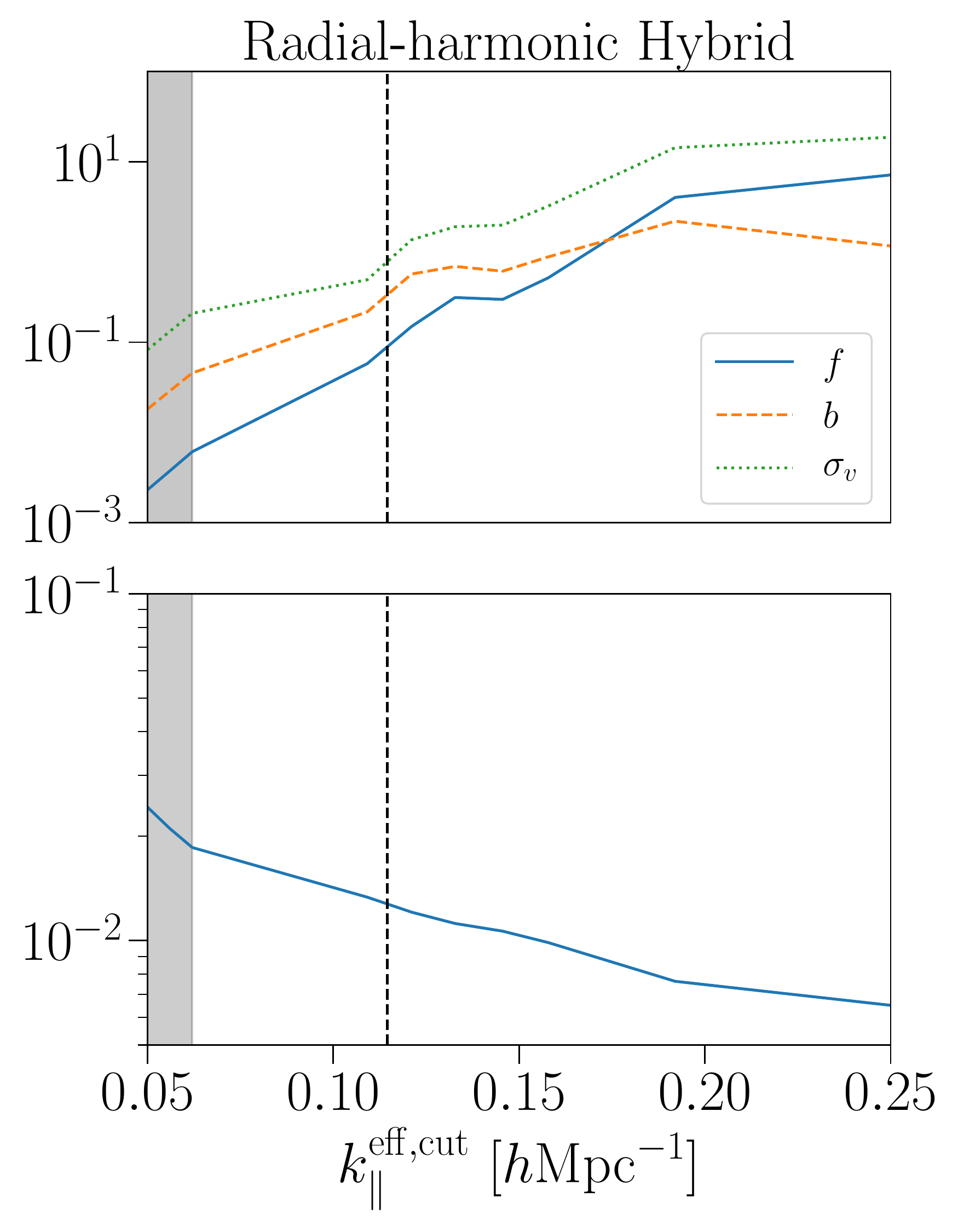}
\caption{Same as Fig.~\ref{fig:results high z}, except the redshift range is now $z \in [1., 1.25]$. The dashed line indicates  the maximum scale cut where the growth function constraints are unbiased in the sense that normalized bias is less than $10 \%$, i.e. $|b| / \sigma < 0.1$. For this choice of scale cut, $\sigma(f)= 0.010$ for the radial-harmonic hybrid estimator, $\sigma(f)= 0.052$ for tomography and $\sigma(f)= 0.011$ for the anisotropic power spectrum.}
\label{fig:results high z}
\end{figure*}

To ensure that our method works over a broad range in redshift, we repeat our analysis for the redshift bin $z \in [1,1.25]$. Assuming the fiducial cosmology, at this redshift $f = 0.89$. As before we assume $f_{\rm sky} = 0.3$ and that the radial window contains 7.5 million galaxies. Assuming $\sigma_v(z)$ scales with the growth factor $D(z)$~\cite{Blanchard:2019oqi} and using our previous fiducial value $\sigma_v(z = 0.675) = 5.1$ implies $\sigma_v(z = 1.125) = 4.2$. We cut perpendicular scales $k_\perp >0.25$, which corresponds to $\ell > 519$ in the projected case. We take $\eta^{\rm min} = 3.4$. This is intended to correspond to the same $k_\parallel$-mode as in the previous section, but we have rescaled $\eta^{\rm  min}$ by the fractional change in the window width following the scaling relation in Eqn.~\ref{eqn:scaling}. 
\par We plot the normalized bias and the error on the growth function in Fig.~\ref{fig:results high z}. As in the previous section, we take the most optimistic scale cut while ensuring that the normalized bias on $f$ does not exceed $10 \%$. For this choice of scale cut, $\sigma(f)= 0.010$ for the radial-harmonic hybrid estimator, $\sigma(f)= 0.011$ for the anisotropic power spectrum and $\sigma(f)= 0.052$ for tomography. Thus the radial-harmonic hybrid estimator again yields an almost fourfold improvement in constraining power compared to tomography.

\section{Conclusion} \label{sec:conclusion}
\par We have presented the formalism for projected anisotropic RSD spectra for an arbitrary radial galaxy weighting in the flat-sky approximation. Writing the projected spectra as a projection of the anisotropic power spectrum over $k_\parallel$-modes weighted by a radial-mode efficiency kernel has provided insights into the efficacy of different radial weighting schemes. 
\par Although commonly applied to photometric survey data, we have argued that the standard projected tomographic power spectra estimator (i.e. top-hat weighting) has several drawbacks when applied to spectroscopic data sets in redshift space. Narrow tomographic binning leads to a mixing of poorly modeled small-scale radial modes with well modelled large-scale radial modes, inducing model bias. Although this can be avoided by choosing wider tomographic bins, even then detailed information about the structure of the anisotropic power spectrum in $(k_\parallel, k_\perp)$-space is lost. This is because for standard tomography the radial efficiency kernels are broad in $k_\parallel$.
\par To avoid mixing between $k_\parallel$-scales, one should choose a weighting which gives narrow radial-mode efficiency kernels in $k_\parallel$-space. One can then also cleanly cut the parts of the data vector which are sensitive to poorly modelled $k_\parallel$-modes. Inspired by the fact that the Fourier transform of a cosine is a delta function, we have argued that a harmonic weighting, labeled by a wavemode-$\eta$, is a natural choice. We refer to the resulting power spectra as the radial-harmonic power spectra.
\par A limiting factor is that for low-$\eta$, the wavelength of the weight function becomes significantly larger than the window. In this regime the harmonic weighting is virtually equivalent to tomography resulting in broad efficiency kernels in $k_\parallel$ and mixing of scales. Hence we advocate using the anisotropic power spectrum to probe large parallel wavemodes and the radial-harmonic $C(\ell)$ to probe small wavemodes, while carefully choosing the scale cuts to ensure the covariance between the two estimators is negligible. 
\par We have performed a Fisher analysis for the three estimators. In the tomographic and radial-harmonic case, scale cuts are defined by the tomographic bin width and $\eta$, respectively. Using the Fisher bias between a Gaussian and Lorentzian FoG as a proxy for model bias due to nonlinear RSD modelling uncertainties we established two key results:
\begin{itemize}
\item When we choose the most optimistic $k_\parallel$-scale cut for each method, while also ensuring the model bias is small, the radial-harmonic hybrid and the anisotropic power spectrum yield significantly tighter constraints on the growth function, $f$, than tomography. 
\item The radial-harmonic hybrid estimator and anisotropic power spectrum estimator yield almost identical constraints showing that it is possible to extract nearly all the information in projection with a suitable radial weighting.
\end{itemize}
\par In an upcoming paper we will extend the radial-harmonic formalism presented in this work to include cross-correlations between weak lensing and RSD and explore the prospects for this type of analysis using current and future data sets.

\section{Acknowledgements}
The authors are grateful for useful conversations with Ruth Durrer, Henry Grasshorn Gebhardt, Mona Jalivand, Tom Kitching and Jason Rhodes and to Anurag Deshpande for providing a Fisher contour plotting routine. AP is a UK Research and Innovation Future Leaders Fellow, grant MR/S016066/1. PLT acknowledges support for this work from a NASA Postdoctoral Program Fellowship. Part of the research was carried out at the Jet Propulsion Laboratory, California Institute of Technology, under a contract with the National Aeronautics and Space Administration. We acknowledge use of the open source software~\cite{2020SciPy-NMeth, 2020NumPy-Array, 4160265}.

\bibliographystyle{apsrev4-1.bst}
\bibliography{bibtex.bib}

\appendix

\section{Numerical Evaluation of the Radial-Mode Efficiency Kernels in the Tomographic Case} \label{sec: tomo app}
\par From Eqn.~\ref{eqn:c1} we see that to evaluate the tomographic power spectra, $C_{ij} (\ell)$, we must integrate the product of the anisotropic power spectrum, $P(k, \mu)$, and the efficiency kernel $\widetilde K_{ij} (k_\parallel)$. Since the power spectrum is naturally expressed in logarithmic space, this implies that to evaluate this integral numerically, we should evaluate $\widetilde K_{ij} (k_\parallel)$ on logarithmic intervals in $k_\parallel$. From Eqn.~\ref{eqn:fourier 1} this reduces to finding the Fourier transform of the tomographic bin window, $W_i(r[z])$, in  log-space. We can not use the Fast Fourier Transform (FFT) as it works on a regular grid.
\par Instead we approximate $W_i(r[z])$ as piecewise function in $z$, and integrate each piece analytically. This is the strategy employed in~\cite{bloomfield2017indefinite} to evaluate integrals over Bessel functions, and may have further applications in cosmological settings where one must rapidly evaluate oscillatory integrals. 
\par We start by approximating $W_i(r[z])$ as a sum of $N$ non-overlapping top-hat functions in $z$. Suppose that $W_i(r[z])$ is only non-zero in the range $[z_{\rm min}, z_{\rm max}]$. Defining $\Delta z = (z_{\rm max} - z_{\rm min}) / N$, $\alpha_n = W_i(r_n)$ where $r_n = r[z_{\rm min} + (n+1/2) \Delta z]$, we can write 
\begin{equation} \label{eqn:piece tomo}
W_i \left( r[z] \right) \approx \sum _{n = 0} ^ {N - 1} \alpha_n T_n (r[z]),
\end{equation}
where $T_n(r[z])$ is a top-hat function 
\begin{equation}
    T_n (r[z])=
    \begin{cases}
      1, & \text{if}\ r_n^- < r[z] < r_n^+ \\
      0, & \text{otherwise}.
    \end{cases}
\end{equation}
Then

\begin{equation} \label{eqn:piece tomo 2}
\begin{aligned}
    \widetilde W_i(k_\parallel) = \sum _{n = 0} ^ {N - 1} \alpha_n \int_{r[z_{\rm min}]}^{r[z_{\rm max}]} {\rm d} r \ T_n\left(r[z] \right) \\ \times \exp \left( -i k_\parallel r[z] \right).
    \end{aligned}
\end{equation}

If we then define $\Delta r_n =  r_n^+ - r_n^-$ where  $r_n = r[z_{\rm min} + (n+1/2) \Delta z]$,  $r^+_n = r[z_{\rm min} + (n+1) \Delta z]$ and $r^-_n = r[z_{\rm min} + n \Delta z]$ we can integrate Eqn.~\ref{eqn:piece tomo 2} analytically~\cite{Gebhardt:2020imr}, so that
\begin{equation}
\widetilde W_i(k_\parallel) =  \sum _{n = 0} ^ {N - 1} \alpha_n \Delta r_{n} e ^{-i k_\parallel r_n} {\rm sinc} \left(\frac{k_\parallel \Delta r_n}{2} \right).
\end{equation}
Hence,
\begin{equation}
\begin{aligned}
    \widetilde K_{ij} (k_\parallel) = \sum_{n,n'} \alpha_n \alpha_{n'} \Delta r_{n} \Delta r_{n'} \cos( k_\parallel[r_{n} - r_{n'}]) \\ \times {\rm sinc} \left(\frac{k_\parallel \Delta r_n}{2} \right) {\rm sinc} \left(\frac{k_\parallel \Delta r_{n'}}{2} \right).
\end{aligned}    
\end{equation}

\section{Numerical Evaluation of the Radial-Mode Efficiency Kernels in the Radial-Harmonic Case} \label{sec:app b}


To evaluate the efficiency kernels in the radial-harmonic case we use the same strategy as in the tomographic case. As in Eqn.~\ref{eqn:piece tomo} we write
\begin{equation}
W(\eta, r[z])  \approx \sum _{n = 0} ^ {N - 1} \alpha_n T_n (r[z]) \cos \left( \frac{2 \pi \eta r^{\rm ref }[z]}{\Delta r^{\rm ref}[z]}\right),
\end{equation}
Hence,
\begin{equation} 
\begin{aligned} \label{eqn:b2}
    \widetilde W(\eta, k_\parallel) = \sum _{n = 0} ^ {N - 1} \alpha_n \int_{r[z_{\rm min}]}^{r[z_{\rm max}]} {\rm d} r \ T_n\left(r[z] \right) \\ \times \cos \left( \frac{2 \pi \eta r^{\rm ref }[z]}{\Delta r^{\rm ref}[z]}\right) \exp \left( -i k_\parallel r[z] \right).
    \end{aligned}
\end{equation}
Now we must write the weight, which is written as a function of the comoving distance in the reference cosmology, as a function of the comoving distance in which we are trying to evaluate the efficiency kernels. Assuming, $\Delta z $, is small we can assume that the comoving distances in the two cosmologies are linearly related inside each piecewise step, $n$, so that
\begin{equation} \label{eq:begin algebra}
r^{\rm ref}[z] = a_n r[z] + b_n,
\end{equation}
To solve for $a_n$ and $b_n$ we first assume than inside each piecewise step the comoving distance is a linear function of redshift
\begin{equation} 
\begin{aligned}
    r[z] &= \nu_n z + \beta_n, \\
    r^{\rm ref}[z] &= \nu^{\rm ref}_n z + \beta_n^{\rm ref}.
\end{aligned}
\end{equation}
A good choice is to take
\begin{equation}
\begin{aligned}
\nu_n &= \frac{r_n^+ - r_n^-}{\Delta z},\\
\nu_n^{\rm ref} &= \frac{r_n^{+,{\rm ref}}  - r_n^{-,{\rm ref} }}{\Delta z}
\end{aligned}
\end{equation}
and 
\begin{equation} \label{eq:end algebra}
    \begin{aligned}
        \beta_n &= r_n, \\
        \beta_n^{\rm ref} &= r_n^{\rm ref},
    \end{aligned}
\end{equation}
where we define $r_n$ etc. as in Section~\ref{sec: tomo app}.
From Eqns.~\ref{eq:begin algebra}-\ref{eq:end algebra} it follows that
\begin{equation}
\begin{aligned}
    a_n &= \frac{\nu_n^{\rm ref}}{\nu_n} \\
    b_n &= \beta_n^{\rm ref} - \beta_n \left( \frac{\nu_n^{\rm ref}}{\nu_n} \right).
    \end{aligned}
\end{equation}
Hence we can write
\begin{equation}
\cos \left( \frac{2 \pi m r^{\rm ref }[z]}{\Delta r^{\rm ref}[z]}\right) = \cos(c_{nm} r[z] + d_{nm})
\end{equation}
where 
\begin{equation}
c_{nm} = \frac{2 \pi m}{\Delta r}, \ d_{nm} = \frac{2 \pi m b_n}{a_n \Delta r}.
\end{equation}
Then using the fact
\begin{equation}
\begin{aligned}
\int ^{b}_{a} {\rm d} x \ \cos (cx +d) e ^{-ikx}  = R[k;a,b,c,d] + iI[k;a,b,c,d]
\end{aligned}
\end{equation}
where
\begin{equation}
\begin{aligned}
R[k;a,b,c,d] &= R2[k;a,c,d] -R1[k;a,c,d] \\&+ R1[k;b,c,d] - R2[k;b,c,d]\\
I[k;a,b,c,d] &= I1[k;a,c,d] + I2[k;a,c,d] \\&- I1[k;b,c,d] - I2[k;b,c,d]
\end{aligned}
\end{equation}
and 
\begin{equation}
\begin{aligned}
    R1[k;r,c,d] &= \frac{c \cos (r k) \sin (r c+d)}{c^{2}-k^{2}},\\
    R2[k;r,c,d] &= \frac{k \sin (r k) \cos (r c+d)}{c^{2}-k^{2}},\\
    I1[k;r,c,d] &= \frac{c \sin (r k) \sin (r c+d)}{c^{2}-k^{2}},\\
    I2[k;r,c,d] &= \frac{k \cos (r k) \cos (r c+d)}{c^{2}-k^{2}}
\end{aligned}
\end{equation}
we can evaluate each piece of Eqn.~\ref{eqn:b2} analytically as in the tomographic case. We find
\begin{equation}
\begin{aligned}
& \widetilde K(k_\parallel; \eta_a, \eta_b)  = \sum_{n, n'} \alpha_n  \alpha_{n'}  \\ &\bigg( R[k_\parallel;r^-_n,r^+_n, c_{n\eta_a}, d_{n\eta_a}]R[k_\parallel;r^-_{n'},r^+_{n'}, c_{n'\eta_b}, d_{n'\eta_b}]  \\&+ I[k_\parallel;r^-_n,r^+_n c_{n\eta_a}, d_{n\eta_a}]I[k_\parallel;r^-_{n'},r^+_{n'} c_{n'\eta_b}, d_{n'\eta_b}] \bigg).
\end{aligned}
\end{equation}

\section{Power Spectrum Estimators in the Radial-Harmonic Basis}
The radial-harmonic $C(\ell)$ can be computed from a spectroscopic galaxy catalogue as follows:
\begin{itemize}
    \item For each $\eta$, compute a pixelized map $m^{\eta}( \boldsymbol{\theta}_P)$. For some pixel, $P$, at an angle $\boldsymbol{\theta}_P$ on the sky, the map is defined as a weighted sum of all galaxies, $g$, inside the pixel. That is
    \begin{equation}
    m^{\eta}( \boldsymbol{\theta}_P) = \sum_{g \in \boldsymbol{\theta_P}} w(\eta, z_g), 
    \end{equation}
    where $z_g$ is the redshift of galaxy $g$ and $w(\eta, z_g)$ is the harmonic weight defined in Eqn.~\ref{eq:weight} assuming a fiducial cosmology to convert redshifts to distances.
    \item Use the map to compute the spherical harmonic coefficients $a_{\ell m} (\eta)$ using e.g. {\tt Healpix}~\cite{Gorski:2004by}.
    \item A good estimate of the pseudo-spectrum (including shot-noise contributions) is then
    \begin{equation} \label{eq:cl data}
    \widetilde C^{\eta_a \eta_b} (\ell) = \frac{w_\ell ^ {-2}}{2 \ell + 1} \sum_{\ell = -m} ^ {m} a_{\ell m}(\eta_a) a^*_{\ell m}(\eta_b),
    \end{equation}
    where $w_\ell$ is the pixel window function which accounts for the suppression of power on small scales due to pixelization~\cite{Leistedt:2013gfa}.
    \item A key feature of the harmonic weight presented in this work is that it is not coupled to angular scales, so that we can use the pseudo-$C(\ell)$ method~\cite{Wandelt:2000av, Alonso:2018jzx} to deconvolve the angular survey mask and estimate $C^{\eta_a \eta_b}(\ell)$. 
\end{itemize}

\section{Configuration Space Radial-Harmonic Formalism and Estimators}
\par Since the radial-harmonic weighting is independent of the angular wavemode, it is possible to compute the radial-harmonic two-point correlations function in configuration space. This sidesteps the need to deconvolve the survey mask which can lead to biases and information loss. If we assume the coupling between $\ell$-modes due to the RSD anisotropy is weak\footnote{This approximation must be tested in a follow-up study.}, we can write the correlation function as a function of the angular power spectrum as in the isotropic case. Then in the flat-sky approximation, the radial-harmonic correlation function, $w^{\eta_a \eta_b} (\theta)$, is given by~\cite{Abbott:2017wau}
\begin{equation}
w^{\eta_a \eta_b} (\theta) = \frac{1}{2 \pi} \int {\rm d} \ell \ \ell J_0(\ell \theta)C ^{\eta_a \eta_b}(\ell),
\end{equation}
where $\theta$ is the angular separation between pairs of galaxies and $J_0$ is the zeroth-order Bessel function. After applying the radial-harmonic weight to the galaxy catalogue, the data vector can be computed using a public code such as {\tt TreeCorr}\footnote{\url{https://github.com/rmjarvis/TreeCorr}}~\cite{Jarvis:2003wq}.

\section{The Likelihood of the Radial-Harmonic Power Spectrum and Hybrid Estimators}
In order to perform a radial-harmonic likelihood analysis, we must know the functional form of the likelihood. In the absence of a mask and assuming that the coupling of $m$-modes due to the RSD anisotropy is weak, each spherical-harmonic coefficient, $a_{\ell m} (\eta_a)$, in the radial-harmonic basis is statistically independent. From Eqn.~\ref{eq:cl  data}, the radial-harmonic power spectrum is a sum over independent, identically distributed random variables, so from the central limit theorem, we expect the likelihood to be Gaussian. For the same reason, binning data vector into bandpowers further Gaussianizes the likelihood. 
\par This Gaussian likelihood approximation must be explicitly checked in a follow-up study. This can be done by performing a mock likelihood analysis using simulated data and comparing to a likelihood-free (e.g.~\cite{Alsing:2019xrx,Akeret:2015uha}) approach as in~\cite{Taylor:2019mgj}. 
\par Meanwhile the likelihood of the anisotropic power spectrum $P(k_\parallel, k_\perp)$ is effectively Gaussian except on extremely large scales where the non-Gaussianity can be handled with existing methods~\cite{Mike:2018zvb}.

\end{document}